\documentclass[12pt]{article}
\usepackage{amsmath} 
\usepackage{graphicx}
\usepackage{subcaption}
\usepackage{float}
\usepackage{placeins}
\usepackage[colorlinks=true, linkcolor=blue, urlcolor=blue, citecolor=blue]{hyperref}
\usepackage[margin=1.0in]{geometry} 
\usepackage{comment}
\usepackage{textgreek}

\newcommand{\tab}{\hspace{0.5cm}}

\begin{document}
\pagenumbering{roman} 
\setcounter{page}{1} 

\title{Simulation of Solar Wind Charged Particle Energy Deposited and Particle Identification by $\Delta$E-E Discrimination in the SNAPPY Cubesat Detector \vspace{1.5cm}}
\author{Daniel Reichart, NASA 2024 Jump Start Student \\Dr. Nickolas Solomey, Prof. of Physics, Advisor \vspace{0.25cm} \\  
\\Wichita State University, Dept. of Mathematics, Statistics and Physics
\\ \\ To be published in the 2024 NASA Jump Start booklet of project summaries}
\date{\today}
\maketitle
\begin{abstract}
    The Solar Neutrino and Astro-Particle PhYsics (SNAPPY) Cubesat is expected to launch in 2025 and it will carry into a polar orbit a prototype test detector for solar neutrino background studies while over the Earth's poles for the neutrino Solar Orbiting Laboratory future project ($\nu$SOL). During this flight it is possible to do other science measurements. One of these is an improved study of the solar wind particles through better particle identification and high resolution energy measurements. This study aimed to understand how well could the solar wind particles be identified using the planned detector but instead of using the veto array as an anti-coincidence it would be used as a $\Delta$E energy sampling of a phoswich particle ID system.
\end{abstract}


\newpage
\hypersetup{linkcolor=black} 
\tableofcontents
\hypersetup{linkcolor=blue} 
\newpage 

\pagenumbering{arabic} 
\setcounter{page}{1} 

\section{Introduction}
$\tab$As a NASA Jump Start undergraduate at Wichita State University, double majoring in Aerospace Engineering and Physics, I am conducting research on the particle discrimination capabilities of the CubeSat detector using the $\Delta$E-E method and measuring the energy spectrum.

The focus of the \textnu SOL project is the detection and study of neutrinos, primarily from the Sun. The research goals of the \textnu SOL project is to utilize the $1/r^2$ neutrino flux with the Sun, determining the neutrino gravitational focus to the Sun, and observing dark matter candidates \cite{Solomey}.

The CubeSat detector design features two different detector materials: Polyvinyltoluene (the Veto) and gadolinium aluminum gallium garnet (called GAGG). Both detector materials are scintillators. The whole detector includes four GAGG crystals that are enclosed within the Veto. The purpose of the GAGG is to measure the neutrino interaction on $^{71}$Ga that is characterized by a double pulse \cite{Solomey}. The purpose of the Veto is to act as a filter for the GAGG. Neutrinos likely will not interact with the Veto's material, and in principle, any particles detected by the Veto should not be neutrinos. Therefore, any double pulse detected in the GAGG that occurs simultaneously with a pulse from the Veto is not considered as a neutrino interaction.

The current plan is for the CubSsat detector prototype to go into low altitude Polar Earth orbit, with an orbital period of 90 minutes, attached to a CubeSat nanosatellite. The detector will be turned on in fifteen minute intervals while the CubeSat is oriented over Earth's north and south poles.

Eventually \textnu SOL plans to go into orbit around the Sun, within a distance of seven solar radii at perihelion. The purpose of the solar orbit is to take advantage of the higher flux of neutrinos at closer distances to the Sun. The overall goal of the \textnu SOL project is to present a new and innovative way of studying the Sun's fusion core and the particle physics of neutrinos. The CubeSat is a test of the detector, but we can use it to study solar wind at Earth and near the Sun in final mission.

\section{\texorpdfstring{Potential Secondary Study: $\Delta$E-E}{Potential Secondary Study: DeltaE-E}}
$\tab$While the main focus of the \textnu SOL project is to detect and study neutrinos, the detector can be used for other research topics. I have been given the task of looking into the cubsat detector's capacity for discriminating solar particles via the $\Delta$E-E method and measuring their energy. This study is to see if and how well the planned detector can measure solar particles as a parasitic mission during or after the neutrino background study.

Generally, \href{https://www.radiation-dosimetry.org/what-is-delta-e-e-detector-telescope-definition/}{$\Delta$E-E telescopes} are composed of two or more detectors placed directly adjacent to one another, such that an incident particle may deposit energy into all detector volumes. The particles of interest generally come from one direction. In $\Delta$E-E studies, data is plotted as the energy deposited in the first detector(s) ($\Delta$E) against the energy deposition in the last detector (E) \cite{Connor}. The $\Delta$E detector(s) are designed to only absorb part of the total energy deposition. The E detector is large enough and dense enough to absorb the remaining particle energy. In some cases, there is a detector after the E detector that, if activated, will cancel the plot for that event, ensuring that particles punching through the E detector are not plotted \cite{Reames}. Particles of different mass deposit different amounts of energy in the detectors, meaning when the data is plotted, particles of different masses show up on different regions in the plot.

The CubeSat particle detector happens to follow a similar design as the $\Delta$E-E telescopes, with some distinct differences.  The GAGG is entirely surrounded by the Veto to ensure neutrino double pulses only occur in the GAGG. Due to this major difference, the CubeSat detector will not follow the standard $\Delta$E-E setup as previously described. The entire Veto acts as the $\Delta$E detector and the GAGG crystals act as the E detector. If a particle has enough energy to punch through both the Veto and the GAGG, energy is again deposited into the Veto which creates deviation from the smooth curves generally associated with $\Delta$E-E studies. Simulation of the CubeSat detector will be required to determine if these deviations are acceptable to justify its use for particle discrimination via $\Delta$E-E.

The best way to go about determining the CubeSat detector's capacity for $\Delta$E-E particle discrimination will be to use Monte Carlo simulations via GEANT4.

\section{GEANT4 Simulation}
$\tab$The software toolkit \href{https://geant4.web.cern.ch/about/}{GEANT4} accurately simulates high energy particle interactions, making the simulations suitable for the interactions necessary to complete the \textnu SOL project $\Delta$E-E particle discrimination study. All of the simulations were performed in GEANT4 to ensure realistic results.

The simulation detector is a modified a model of a detector that had already been built in GEANT4 \cite{Doty}. The modified detector only required additional shielding to the outer layer of the detector volumes and minor adjustments to the dimensions. Implementing scenarios with realistic particles and energy ranges were the only significant changes necessary.

\subsection{Geometry}
$\tab$While the geometry of the CubeSat detector is not overly complex, it includes many volumes that are important to the functionality of the detector as a whole. 

\begin{figure}[h!]
    \centering
    \begin{minipage}[b]{0.45\textwidth}
        \centering
        \includegraphics[height=4.8cm]{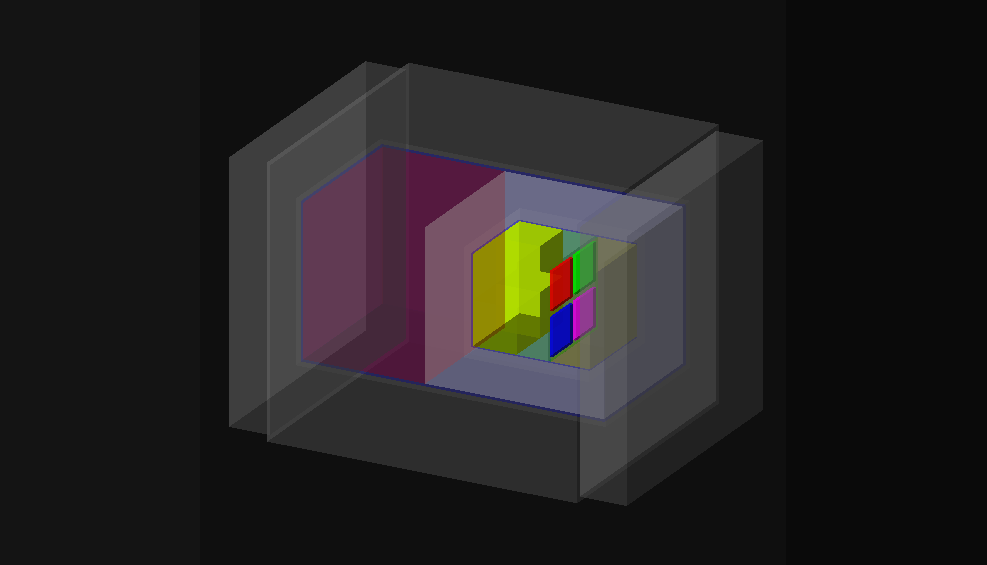}
        \caption{The CubeSat detector overview.}
        \label{Overview}
    \end{minipage}
    \hspace{0.05\textwidth} 
    \begin{minipage}[b]{0.45\textwidth}
        \centering
        \includegraphics[height=4.8cm]{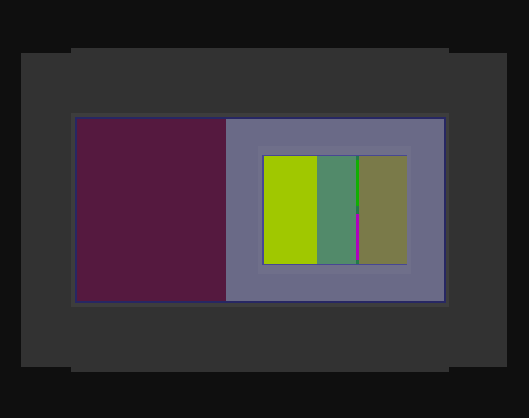}
        \caption{The CubeSat detector side view.}
        \label{Outline}
    \end{minipage}
\end{figure}

The gray volumes around the outside is the tungsten-epoxy shielding \cite{Novak}. The purple volume is the Veto PMT, which collects the photons that are produced in the Veto. The light blue volume is the Veto. The light brown volume is the electronics for the GAGG. The green volume is a quartz crystal window. The yellow volume is the four GAGG crystals. There is an air gap between the shielding and the main detector volumes that is portrayed in light gray. The detector also has an aluminum coating surrounding the main detector volumes and the inner Veto volumes shown in very thin, dark blue shading. In Figure~\ref{Overview}, the vibrant red, green, blue, and pink volumes are the SiPMs for the GAGG crystals. The SiPMs collect the photons produced in the GAGG. The GAGG payload refers to the GAGG, quartz crystal window, and electronics for the GAGG. Epoxy fills the gap between the GAGG payload and the Veto to prevent movement. In the simulation, this volume is the same color as the air gap between the main detector volumes and the shielding. Additional specifics of the CubeSat detector design can be viewed in \textit{Design and Testing of a 3U CubeSat to Test the In-situ Vetoing for the \textnu SOL Solar Neutrino Detector} \cite{Folkerts}. Figure~\ref{Overview} shows a complete assembly of the CubeSat detector. For the simplicity of the simulation, the CubeSat containing the detector was not built. While the effects of the shielding provided by the CubeSat is small, it is not insignificant and may be the focus of further study.

To accurately simulate the true shielding density, the GEANT4 program incorporates the epoxy-to-tungsten powder ratio based on the materials' known densities ($\rho$) and masses (m) used to create the shielding. The density calculation utilizes the well know formula:

\begin{equation*}
    \rho_{\text{shielding}} = \frac{m_{\text{Total}}}{V_{\text{Total}}} = \frac{m_{\text{W}} + m_{\text{E}}}{V_{\text{W}} + V_{\text{E}}} = \frac{m_{\text{W}} + m_{\text{E}}}{\frac{m_{\text{W}}}{\rho_{\text{W}}} + \frac{m_{\text{E}}}{\rho_{\text{E}}}}
\end{equation*}


The hope of outlining the CubeSat detector is to visualize the differences between this detector and the general $\Delta$E-E telescope. The GAGG (E detector) is completely encased in the Veto ($\Delta$E detector). Based on the differences in geometry, there will be deviations from the expected results of usual $\Delta$E-E studies.

\subsection{Detectors}
$\tab$The primary detection method of the CubeSat detector is scintillation light. Scintillators produce scintillation light due to passage of charged particles. There are five detectors that work in tandem within the GEANT4 CubeSat detector simulation. The Veto's photomultiplier tube (PMT) reads out the number of photons that are produced in the Veto. The four SiPMs read out the number of photons produced in the four corresponding GAGG crystals. For simulation purposes, the photomultipliers count the number of photons produced in their respective scintillators. In reality, the photomultipliers in the CubeSat detector do not count the number of photons, instead reading out a voltage that is directly proportional to the number of photons that are produced. This voltage will correspond to the amount of energy that is deposited in that scintillator volume.

The ability of a scintillator to produce scintillation light is dependent on the material properties of the scintillator in use. Properties such as index of refraction, scintillation yield, resolution scale, and absorption length are important material properties to define in GEANT4 to accurately simulate scintillation light. For this study, I am using the same material properties that were defined in the given GEANT4 model, with the exception of scintillation yield. The length of time required to complete a run when the scintillation yield is at the true value is excessive. To combat this issue, I am running scintillation yield at exactly 1\% of the true material value. Though this value of scintillation yield is lower, it should not affect the overall shape of the graphs. In practice, the shape of the plots will be proportional to the simulation plots, with the only notable difference being that the voltage is read instead of the photon count.

\subsection{Particle Specifics}
$\tab$An important part of this study has been determining what particles can be expected to get through the shielding, and what energy ranges those particles may have. Particles on solar wind generally do not have enough kinetic energy to get through the shielding. However, the Sun does produce solar energetic particles. Solar energetic particles can range anywhere between 100 keV to hundreds of MeV or a couple of GeV. The particle composition can be as lightweight as electrons to as heavy as Iron ions. With this in mind, not all particles in solar energetic particle events will have the same kinetic energy. The kinetic energy of solar energetic particles tend to vary with mass.

In order to realistically simulate solar energetic particles with the CubeSat detector, I need to determine: which particles have the highest probability of being on solar energetic particle events, which particles will get through the shielding, and what energies those particles can have. Worth noting, there are over 30 different particles and ions that solar energetic particles are known to include. The simulation will start with a lower number of particles, and increase as necessary.

In this simulation, the particle and ion types that were chosen with information found in \cite{Reames}. The elemental isotopes were determined with information found in \cite{Leske}. For the simplicity of the simulation, none of the ions will carry any bound electrons.

To determine the upper energy ranges that a particle may realistically have, I will enforce a constraint on the differential particle flux, or intensity. For this study I set this minimum differential particle flux to $10^{-4}\text{ }\frac{\text{particles}}{\text{cm}^2\text{ * s * sr}}$, to limit the high energy particles to ones that we may actually detect. For ease of coding, the lowest energy (per nucleon) that any particle will have in this simulation is 1 MeV. Worth noting, the simulation is currently assigning particles their energy on a logarithmic distribution that favors lower energies. This is a good approximation, but is not the same as observation.

The numbers that are provided in this section were obtained using PlotDigitizer, a pixel counting software, to analyze the plots from the sources mentioned below. PlotDigitizer is used to estimate the spectra of a particle flux of $10^{-4}\text{ }\frac{\text{particles}}{\text{cm}^2\text{ * s * sr}}$. Nothing below this differential particle flux is used in the simulation.

The energy range for electrons was determined from graphs found in \cite{McGuire}. The energy spectra of protons, alpha particles, and $^3$He were found in \cite{Reames}. The remaining energy spectra are deduced from \cite{Desai}. 

Table 1 summarizes the particles/ions and associated energies that were determined from the sources mentioned above. These energy ranges, while approximate, are a good model for solar energetic particle activity.

\begin{table}[h!]
    \centering
    \label{Table}
    \renewcommand{\arraystretch}{1.5}
    \begin{tabular}{|c|c|}
        \hline
        Particles/Ions & Energy Ranges \\
        \hline
        e$^-$   & 1 MeV - 10.15 MeV   \\
        \hline
        $^1_1$H$^{+}$   & 1 MeV - 2.55 GeV   \\
        \hline
        $^3_2$He$^{+2}$   & 3 MeV - 19.50 MeV   \\
        \hline
        $^4_2$He$^{+2}$   & 4 MeV - 3.16 GeV   \\
        \hline
        $^{12}_6$C$^{+6}$   & 12 MeV - 1.95 GeV   \\
        \hline
        $^{14}_7$N$^{+7}$   & 14 MeV - 1.99 GeV   \\
        \hline
        $^{16}_8$O$^{+8}$   & 16 MeV - 2.08 GeV   \\
        \hline
        $^{20}_{10}$Ne$^{+10}$   & 20 MeV - 1.19 GeV   \\
        \hline
        $^{24}_{12}$Mg$^{+12}$   & 24 MeV - 1.94 GeV   \\
        \hline
        $^{28}_{14}$Si$^{+14}$   & 28 MeV - 1.94 GeV   \\
        \hline
        $^{56}_{26}$Fe$^{+26}$   & 56 MeV - 1.58 MeV   \\
        \hline
    \end{tabular}
    \caption{Expected particles and ions along with their energy ranges.}
\end{table}

\subsection{Simulation Scenarios}
\begin{figure}[h!]
    \centering
    \includegraphics[width=0.64\textwidth]{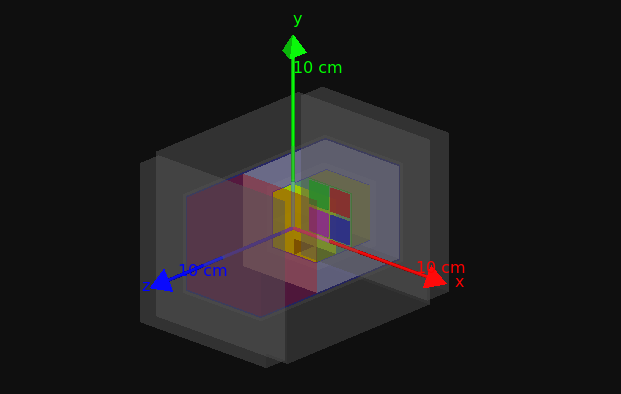}
    \caption{Positive directions as specified by the coordinate axis system in GEANT4.}
    \label{Directions}
\end{figure}

$\tab$This section will explain the seven different simulation scenarios that I have developed over the course of this study. In order to do this, a coordinate axis system needs to be defined. Luckily, GEANT4 already has this covered. In Figure~\ref{Directions}, the red arrow shows the positive X direction, the green arrow shows the positive Y direction, and the blue arrow shows the positive Z direction. This axis system is necessary to understand the different faces of the CubeSat detector that will be under the microscope in these scenarios.

\begin{figure}[h!]
    \centering
    \begin{subfigure}[b]{0.3\textwidth}
        \centering
        \includegraphics[width=\textwidth]{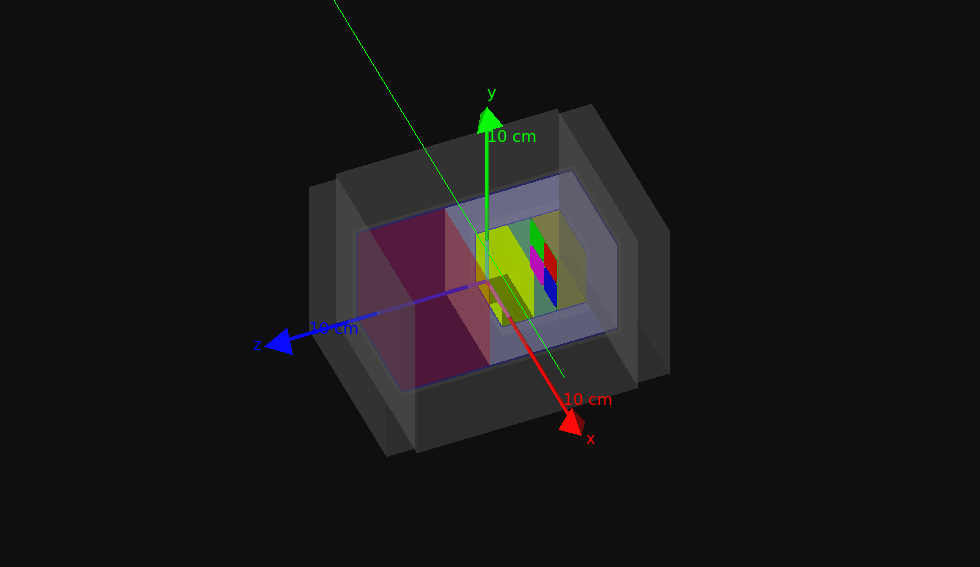}
        \caption{Pencil Beam goes through the Veto and the GAGG -- pencil beam}
        \label{PencilSim}
    \end{subfigure}
    \hfill
    \begin{subfigure}[b]{0.3\textwidth}
        \centering
        \includegraphics[width=\textwidth]{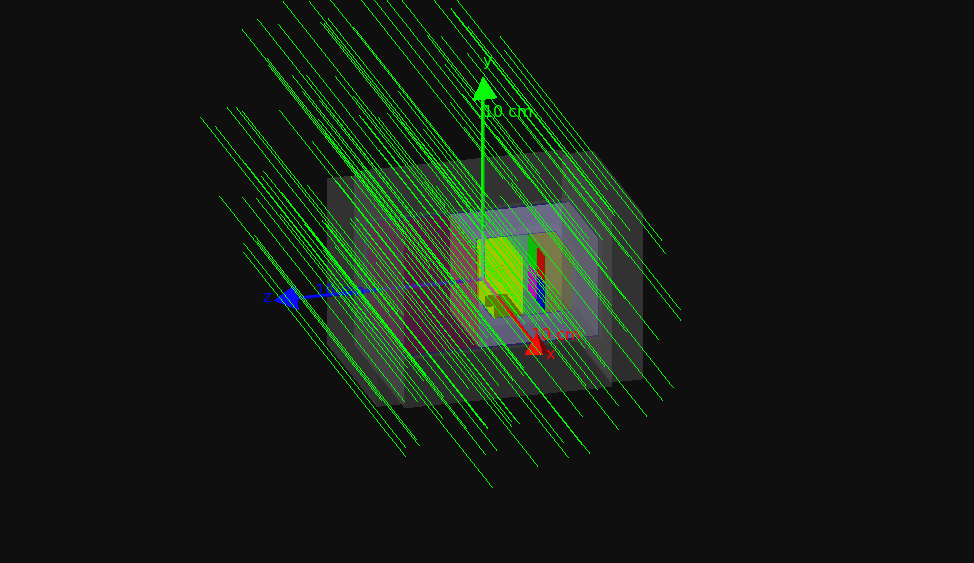}
        \caption{The Particles Attack the X(Y) Face of the Detector -- xyFace}
        \label{xyFaceSim}
    \end{subfigure}
    \hfill
    \begin{subfigure}[b]{0.3\textwidth}
        \centering
        \includegraphics[width=\textwidth]{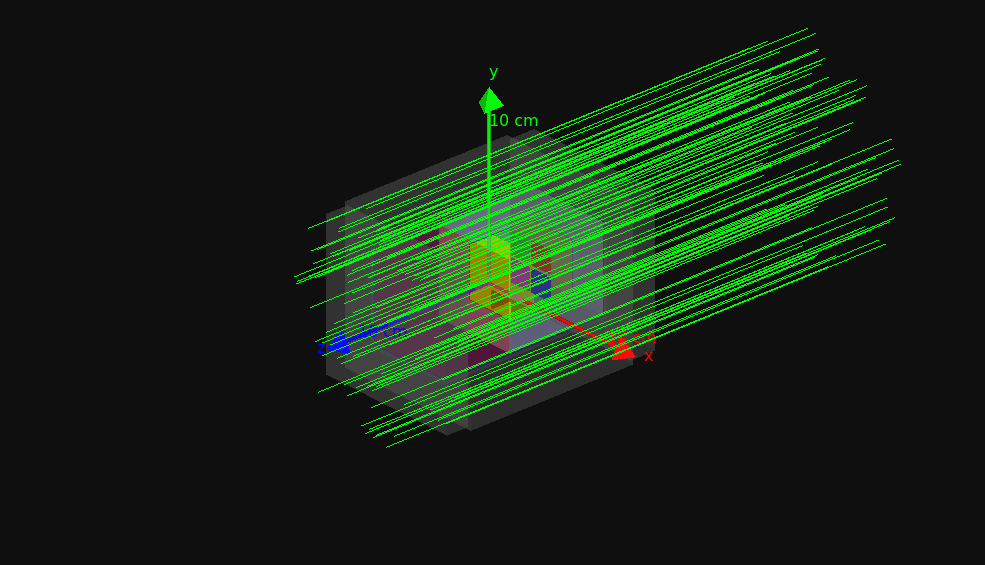}
        \caption{The Particles Attack the +Z Face of the Detector -- z+Face}
        \label{z+FaceSim}
    \end{subfigure}    
    \vspace{1em} 
    \begin{subfigure}[b]{0.3\textwidth}
        \centering
        \includegraphics[width=\textwidth]{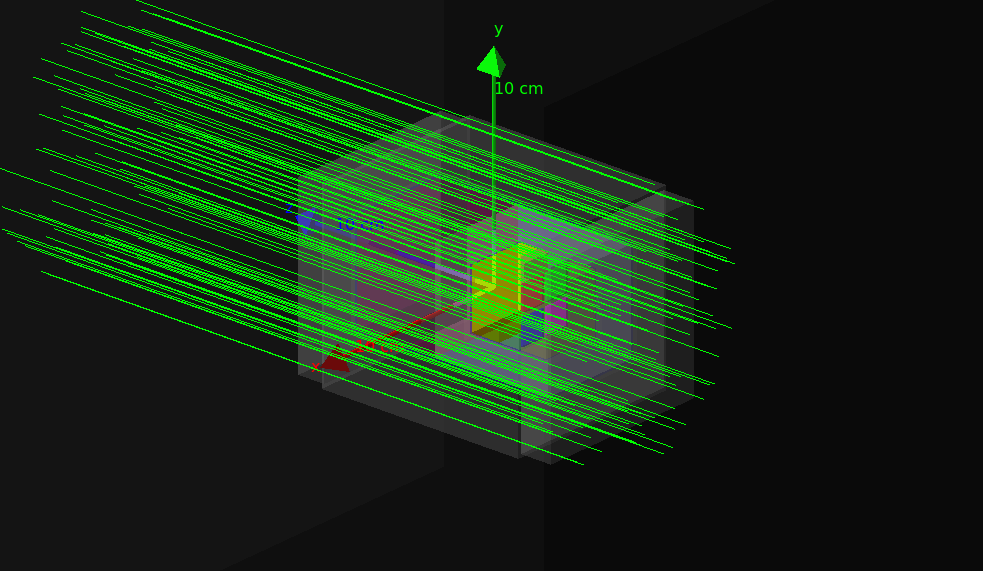}
        \caption{The Particles Attack the -Z Face of the Detector -- $\text{z-Face}$}
        \label{z-FaceSim}
    \end{subfigure}
    \hfill
    \begin{subfigure}[b]{0.3\textwidth}
        \centering
        \includegraphics[width=\textwidth]{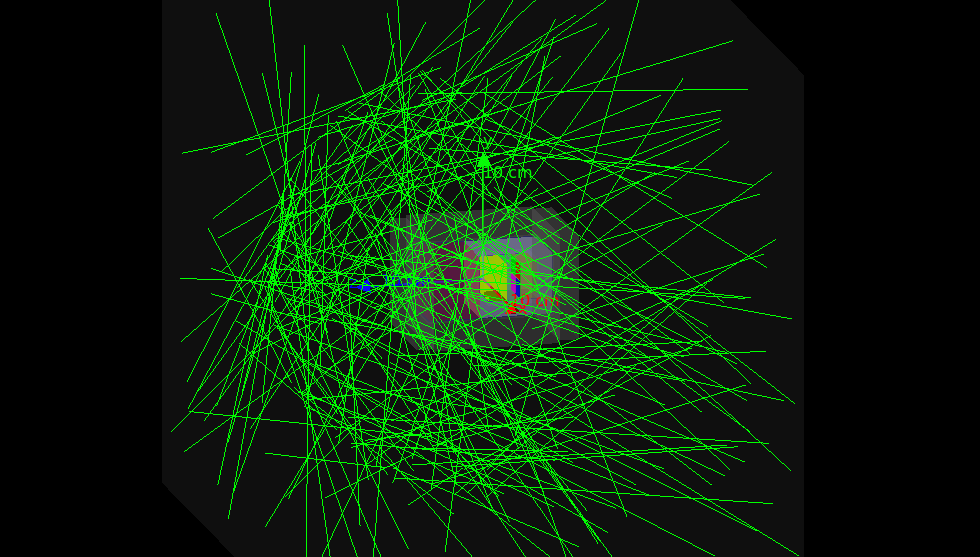}
        \caption{Hemisphere Isotropic Distribution for the +Z Face -- hemisphereIsoZ+Face}
        \label{IsoSim}
    \end{subfigure}
    \hfill
    \begin{subfigure}[b]{0.3\textwidth}
        \centering
        \includegraphics[width=\textwidth]{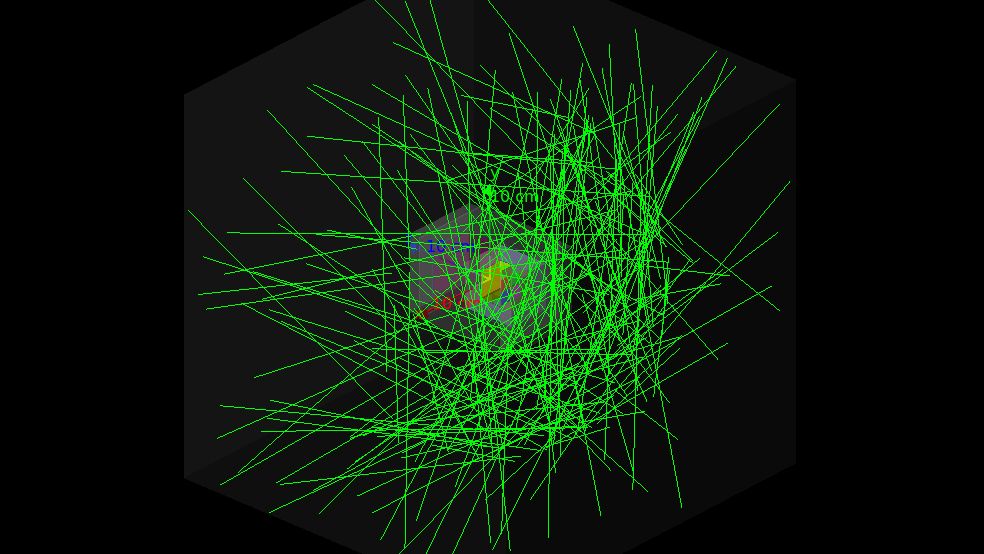}
        \caption{Hemisphere Isotropic Distribution for the -Z Face -- hemisphereIsoZ-Face}
        \label{HemisphereIsoZ+FaceSim}
    \end{subfigure}
    \vspace{1em} 
    \begin{subfigure}[b]{0.3\textwidth}
        \centering
        \includegraphics[width=\textwidth]{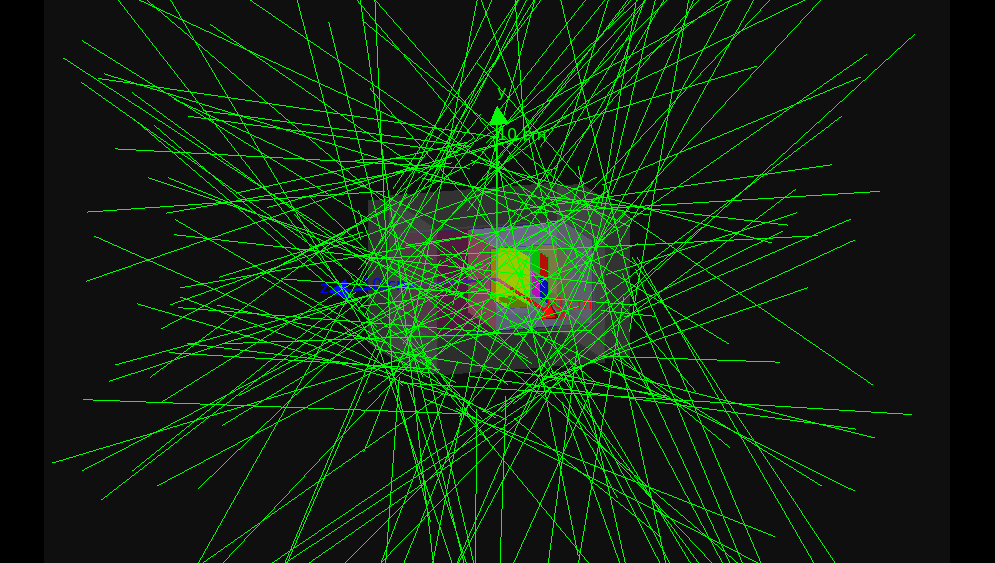}
        \caption{Spherical Isotropic Distribution -- iso}
        \label{HemisphereIsoZ-FaceSim}
    \end{subfigure}
    \caption{The seven simulation scenarios that are implemented in GEANT4. }
\end{figure}

The first simulation that I implemented was a pencil beam. The pencil beam will go through the +X face of the shielding, and directly through the Veto and GAGG. This scenario will not be encountered by the CubeSat detector, but will provide a reference as to what sort of results can be expected at ideal conditions. 

The next step up from the pencil beam is sending the particles at the entire detector face. The X faces and Y faces are geometrically the same, and so the scenario in which these faces are the focus is the xyFace scenario. Other important faces to attack are the +Z face and -Z face, and their scenarios are called z+Face and z-Face respectively. These scenarios will provide a reference for the results that can be expected from near ideal conditions, but these scenarios will most likely not be encountered by the CubeSat detector while in low Earth orbit.

In low Earth orbit, the particle distribution on the detector is expected to be isotropic. To simulate this, the last three scenarios concern an isotropic distribution of the particles. The first is a spherical isotropic distribution, with the particles being initialized on the surface of a sphere surrounding the detector and then shot isotropically at the detector. This scenario will be called iso. The second and third are both hemispherical isotropic distributions, with one being directed at the +Z face and the other being directed at the -Z face. The purpose of these last two scenarios are to simulate the detector traveling over the North and South poles while in low Earth orbit. These scenarios will be called hemisphereIsoZ+Face and hemisphereIsoZ-Face.

These scenarios should provide a good evaluation of the CubeSat detector's capacity for $\Delta$E-E particle discrimination. The Face scenarios will give a good one to one comparison between general $\Delta$E-E telescopes and the CubeSat detector. The Isotropic scenarios will provide us with a good look at how well the CubeSat detector can discriminate particles given the environment it will be in. 

\section{Results}
$\tab$The differences that will be found between the energy deposit and photon count graphs are due to a multitude of reasons including geometry and optical transport. The reasoning of these differences will not be discussed in this report. It will instead focus on how these differences can help differentiate between the simulation scenarios, and various particle types.

The energy deposition that GEANT4 reads is essentially perfect information. No detector, given an imperfect quantum efficiency, will read out the same energy deposits as GEANT4. Thus, a more accurate depiction of energy deposition is the signals that the detectors will read. Photon count serves this purpose. However, the best way to validate the results of photon count plots is to compare it against the energy deposition plots. There will inevitably be some differences, however the general particle regions should remain separate.

The following graphs were constructed using the GEANT4 simulation data with the data handling and analysis software \href{https://root.cern/about/}{ROOT}, developed by CERN. 

\subsection{Energy Deposition vs Total Particle Energy Graphs}

\begin{figure}[h!]
    \centering
    \begin{subfigure}[b]{0.44\textwidth}
        \centering
        \includegraphics[width=\textwidth]{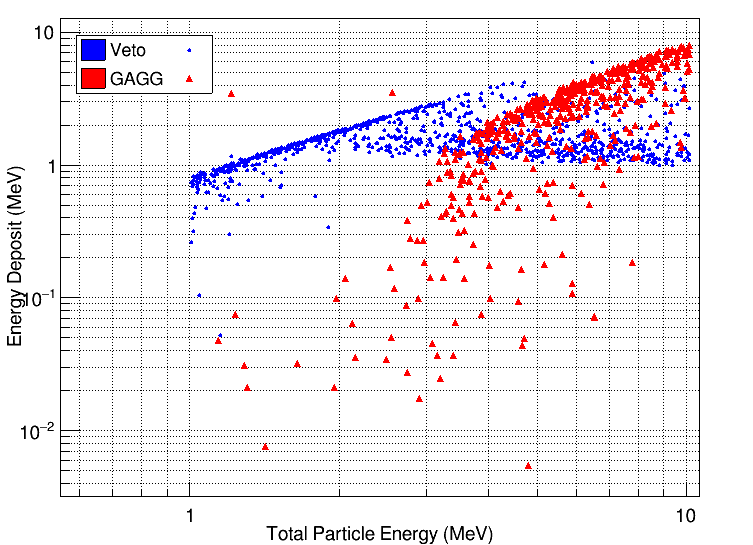}
        \caption{Electrons}
        \label{electron}
    \end{subfigure}
    \hfill
    \begin{subfigure}[b]{0.44\textwidth}
        \centering
        \includegraphics[width=\textwidth]{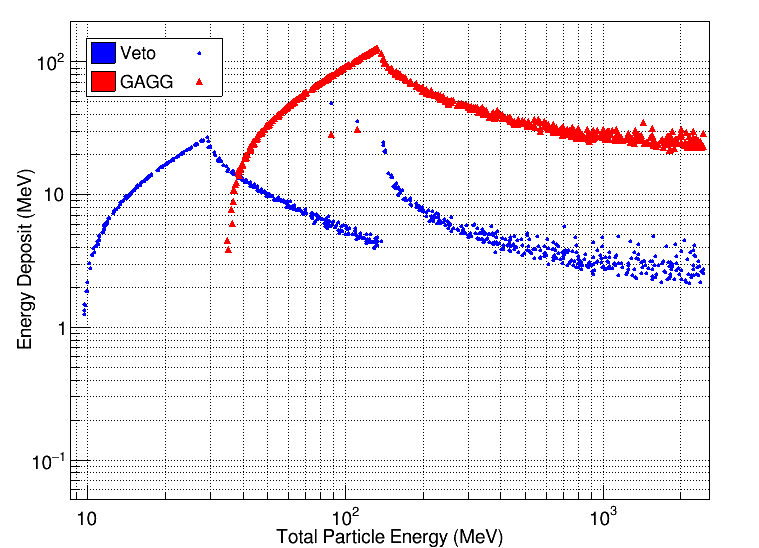}
        \caption{Protons}
        \label{proton}
    \end{subfigure}
    \begin{subfigure}[b]{0.44\textwidth}
        \centering
        \includegraphics[width=\textwidth]{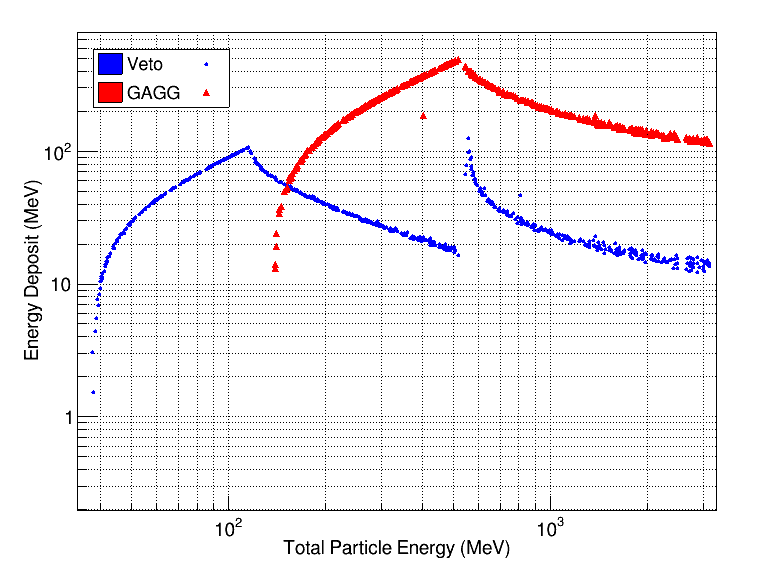}
        \caption{Alpha Particles}
        \label{alpha}
    \end{subfigure}    
    \caption{Electrons show chaos while protons and alphas exhibit discontinuities.}
    \label{EPA}
\end{figure}

\begin{figure}[h!]
    \centering
    \begin{subfigure}[b]{0.45\textwidth}
        \centering
        \includegraphics[width=\textwidth]{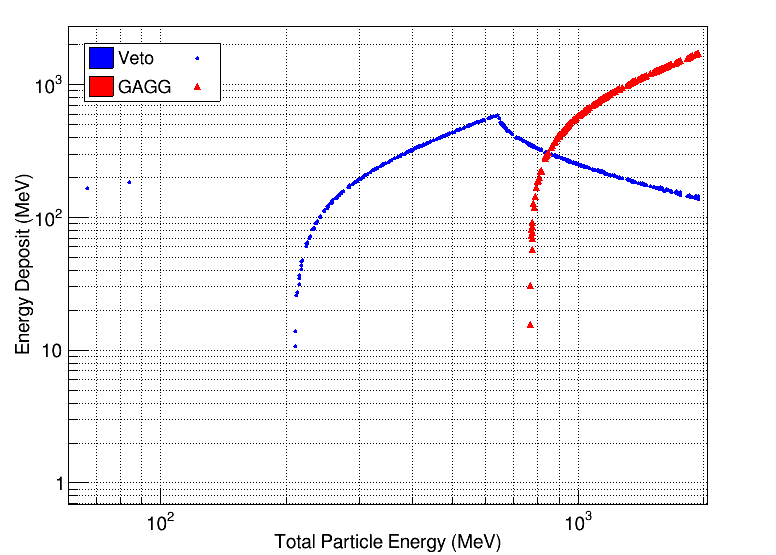}
        \caption{Carbon}
        \label{carbon}
    \end{subfigure}
    \hfill
    \begin{subfigure}[b]{0.43\textwidth}
        \centering
        \includegraphics[width=\textwidth]{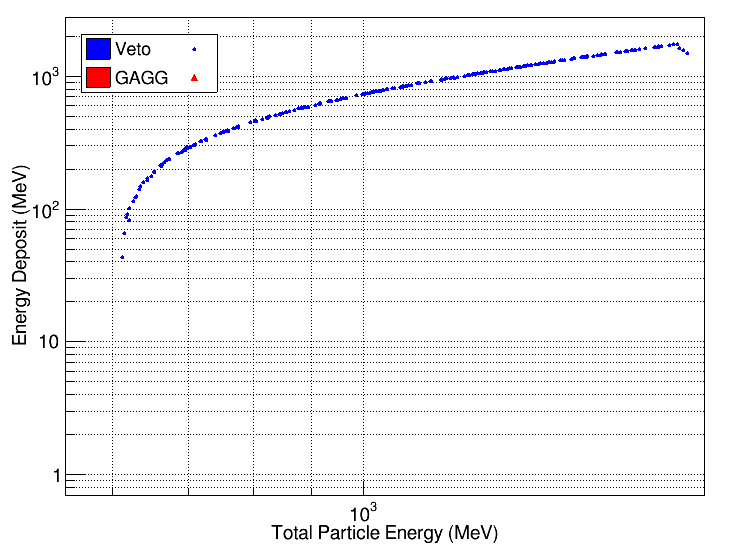}
        \caption{Magnesium}
        \label{magnesium}
    \end{subfigure}   
    \caption{Carbon shows no discontinuities, while Magnesium has no GAGG signal.}
    \label{notEPA}
\end{figure}

$\tab$These plots showcase which particles have the capability of getting through the detector shielding. In Figure~\ref{EPA} and Figure~\ref{notEPA}, only five graphs are shown. $^3$He and $^{56}$Fe are not presented here because all of the particle energy is lost within the shielding, and therefore do not deposit any energy in the Veto or the GAGG. Nitrogen, Oxygen, and Neon are not shown because they are very similar to Figure~\ref{carbon}, while Silicon is not show shown because it is very similar to Figure~\ref{magnesium}. In Figure~\ref{EPA} and Figure~\ref{notEPA}, the blue shows energy deposited in the Veto and the red shows the energy deposited in the GAGG. These plots were taken from the ideal pencil beam scenario.

The pencil beam initializes the particles very close to the detector, with all the particles aimed to go through the Veto and GAGG from the shortest distance. The other scenarios will be less uniform compared to the pencil beam, which represents the ideal scenario. Therefore any particle that is not detected in the ideal scenario, should not be detected in any of the other scenarios.

In Figure~\ref{EPA}, the plots b \& c showcase what it will look like when particles deposit energy into the Veto twice. This abrupt change in the Veto energy deposit will show up later in the $\Delta$E-E plots, but for clarity I would like to address it here. The $\Delta$E-E plots are the energy deposit (or photon count) in the Veto plotted against the energy deposit (or photon count) in the GAGG. The plots for protons and alpha particles will have discontinuities due to the energy deposition jump in the Veto at the punch-through energy of the GAGG.

The plots in Figure~\ref{notEPA} tell us that $\Delta$E-E will not be valid for some of the chosen simulation particles. $\Delta$E-E measurements can only be taken if the incident particle deposits energy in both of the detectors. Therefore, it is expected that the CubeSat detector will not be capable of particle discrimination for $^3$He, $^{24}$Mg, $^{28}$Si, or $^{56}$Fe within the scenarios and energy ranges considered in this simulation.

\subsection{Pencil Beam}
\begin{figure}[h!]
    \centering
    \includegraphics[width=0.9\textwidth]{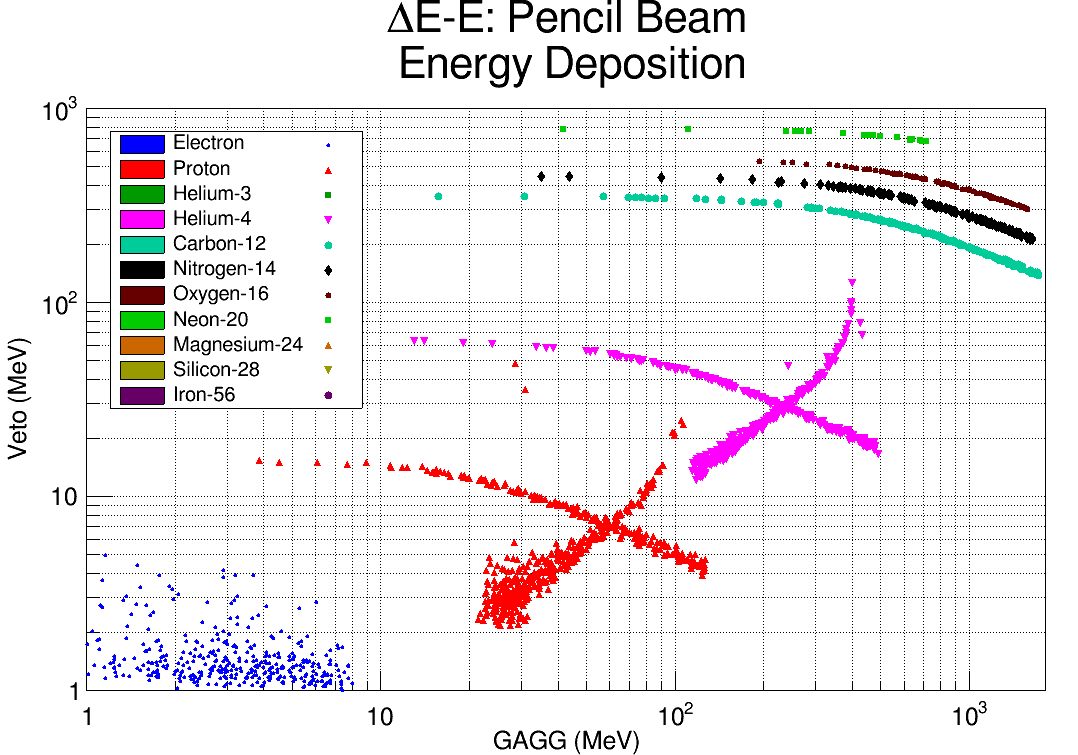}
    \caption{Veto vs GAGG energy deposition particle plot for the Pencil Beam scenario.}
    \label{pencilEdepP}
\end{figure}

\begin{figure}[h!]
    \centering
    \includegraphics[width=0.82\textwidth]{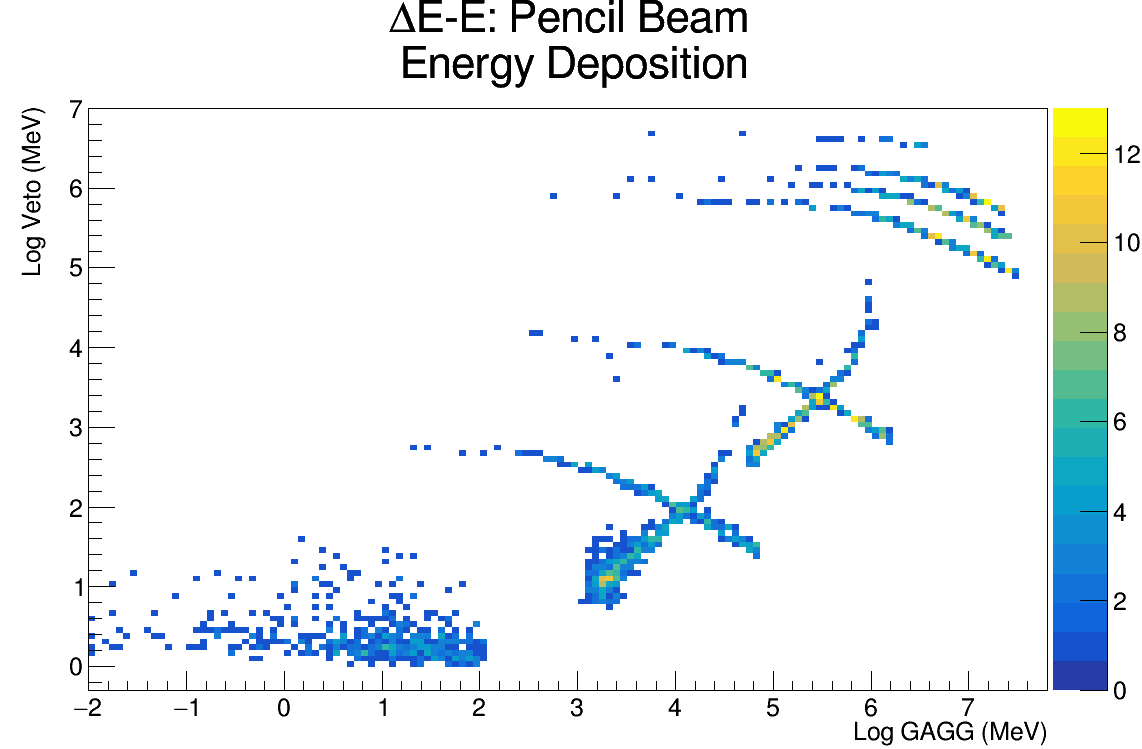}
    \caption{Veto vs GAGG energy deposition histogram for the Pencil Beam scenario.}
    \label{pencilEdepH}
\end{figure}

The purpose of Figure~\ref{pencilEdepP}, and subsequent particle plots, is to visualize what each region of the histogram represents. When there is a signal in the Veto and GAGG, it will show up on the plot without a label, as shown by the histogram in Figure~\ref{pencilEdepH}. Figure~\ref{pencilEdepP} shows what particles take up the different regions of the histogram. This plot can also be a visual representation of what particles got through the shielding on that particular simulation. In Figure~\ref{pencilEdepP} there are no dark green, orange, gold, or purple dots, indicating that no $^3$He, $^{24}$Mg, $^{28}$Si, or $^{56}$Fe deposited energy in both the Veto and GAGG.

The histogram in Figure~\ref{pencilEdepH} will be more accurate to what will be read by the detector during its operation. Knowing which regions correspond to each particle is necessary information for particle discrimination. 

\begin{figure}[h!]
    \centering
    \includegraphics[width=0.82\textwidth]{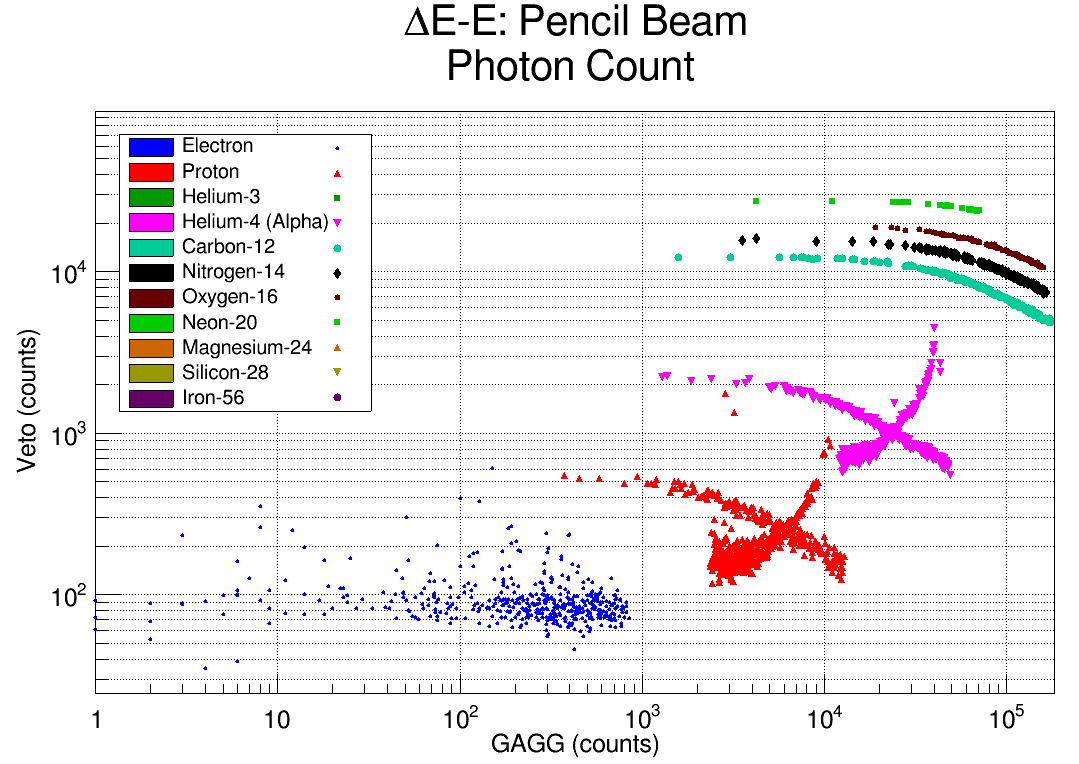}
    \caption{Veto vs GAGG photon count particle plot for the Pencil Beam scenario.}
    \label{pencilPhotonP}
\end{figure}

\begin{figure}[h!]
    \centering
    \includegraphics[width=0.9\textwidth]{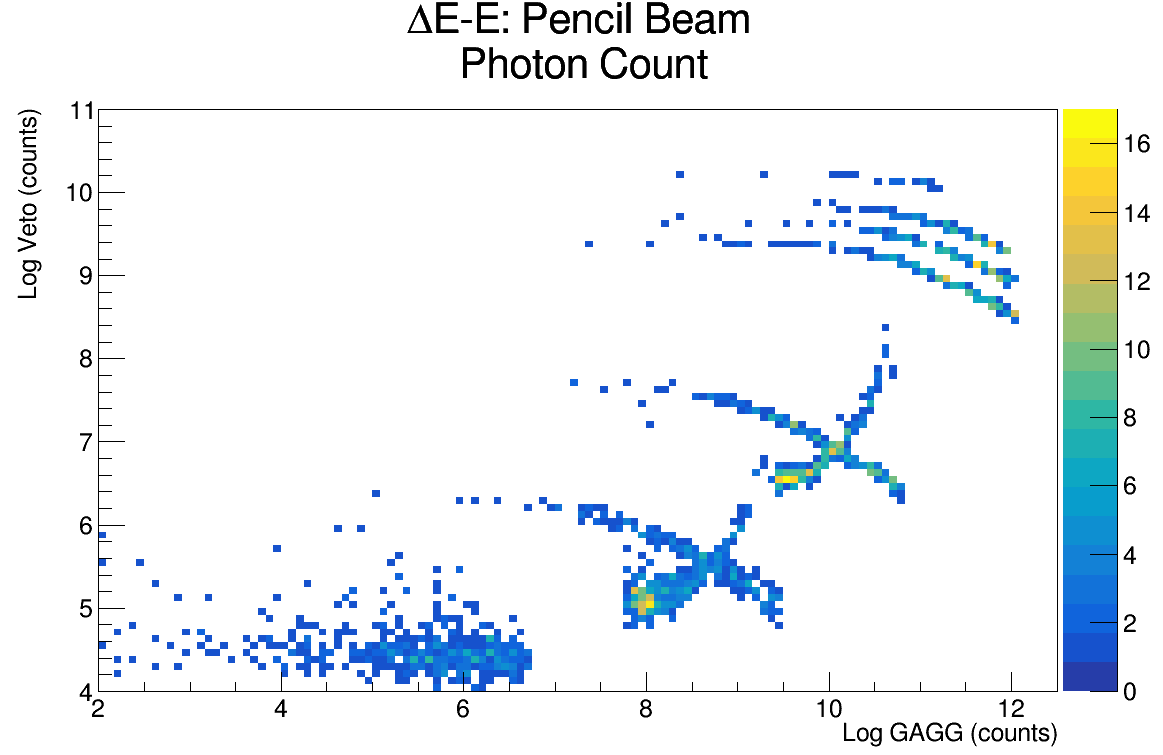}
    \caption{Veto vs GAGG photon count histogram for the Pencil Beam scenario.}
    \label{pencilPhotonH}
\end{figure}

Looking at the results of the photon count study, the expected result should be very similar to the results of the energy deposition study, and as can be seen in Figure~\ref{pencilPhotonP} and Figure~\ref{pencilPhotonH}, the results of the photon count study are very similar to the results found by the energy deposition study. There is a slight variation in the photon count study but, as stated before, this report will not be analyzing why these differences occur. 

\subsection{Face Simulations}
$\tab$The face simulations correspond to the simulation scenarios in which the all the particles in the simulation come from the same direction, attacking the entire face in question. As stated before, these scenarios are called xyFace, z+Face, and z-Face.

\subsubsection{Impingement of the xyFace}
\begin{figure}[h!]
    \centering
    \includegraphics[width=0.88\textwidth]{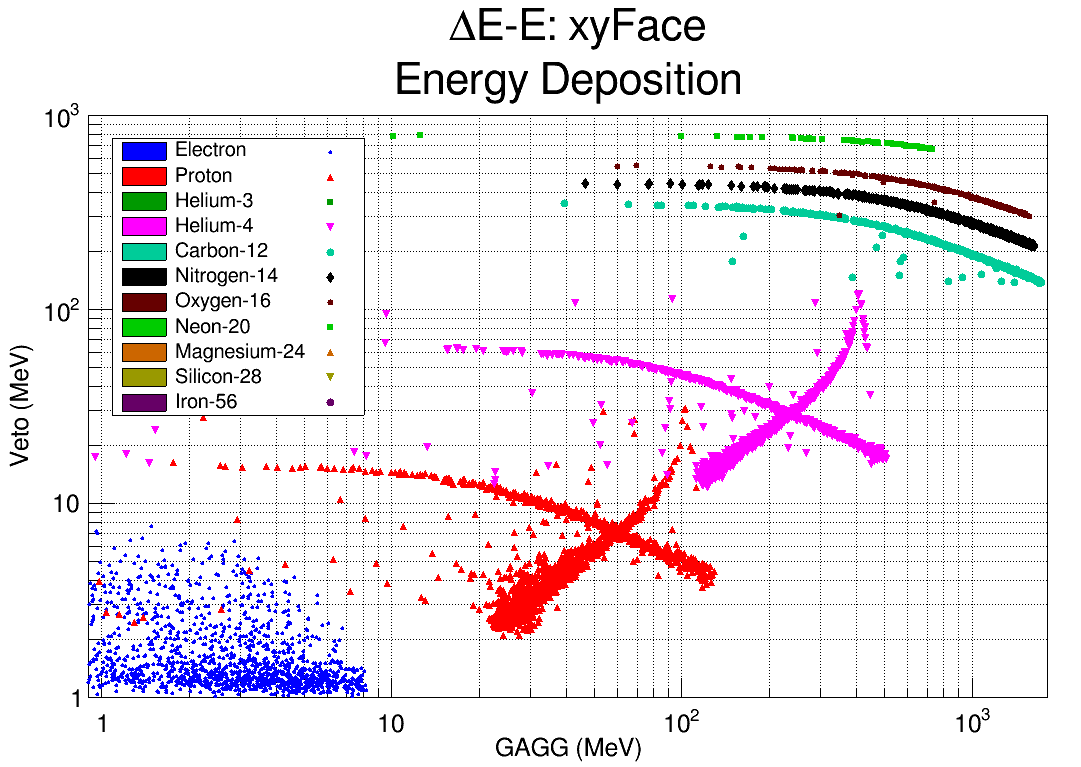}
    \caption{Veto vs GAGG energy deposition particle plot for the xyFace scenario.}
    \label{xyFaceEdepP}
\end{figure}

\begin{figure}[h!]
    \centering
    \includegraphics[width=0.88\textwidth]{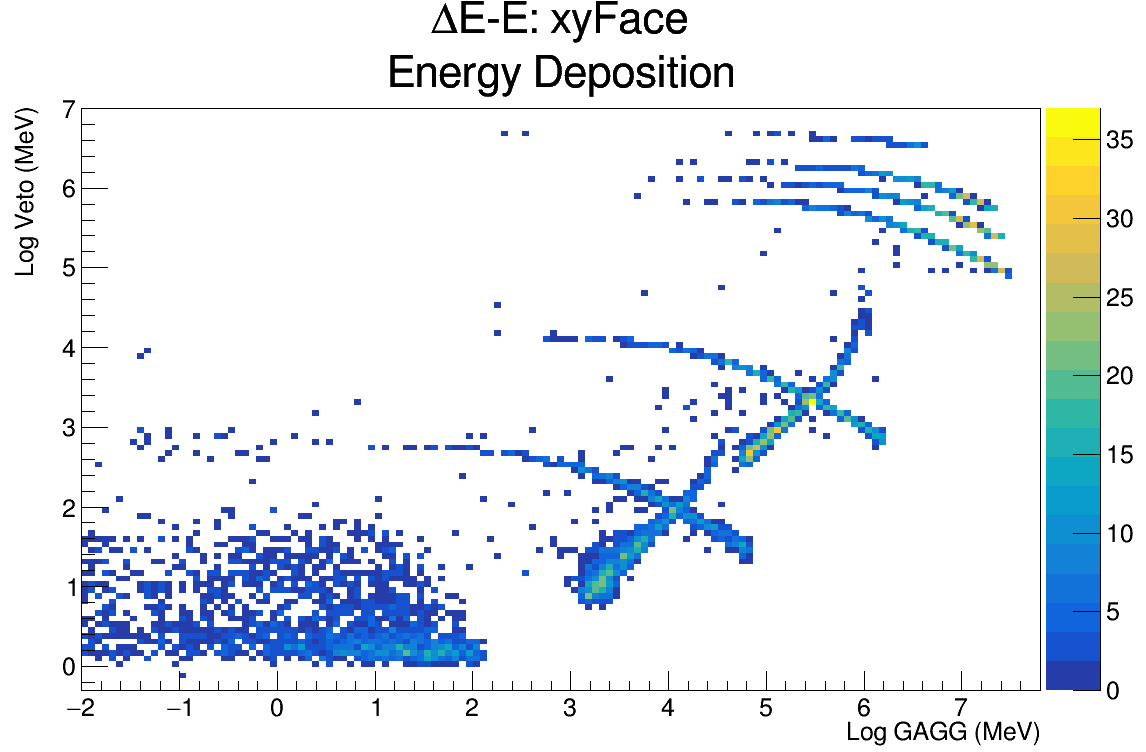}
    \caption{Veto vs GAGG energy deposition histogram for the xyFace scenario.}
    \label{xyFaceEdepH}
\end{figure}

$\tab$Beginning with the xyFace scenario, it can be seen from Figure~\ref{xyFaceEdepP} and Figure~\ref{xyFaceEdepH} that the energy deposition for the xyFace is very similar to that of the pencil beam. This can be explained by their similar nature, with both scenarios attacking the Veto and GAGG from the same directions. It is also found that the same particles that get through the shielding in the pencil beam scenario also get through the shielding in the xyFace scenario. 

It can be seen that the xyFace scenario deviates slightly from the pencil beam scenario. This is due to the distribution of particles over the entire face of the detector, and not just a single point.

\begin{figure}[h!]
    \centering
    \includegraphics[width=0.88\textwidth]{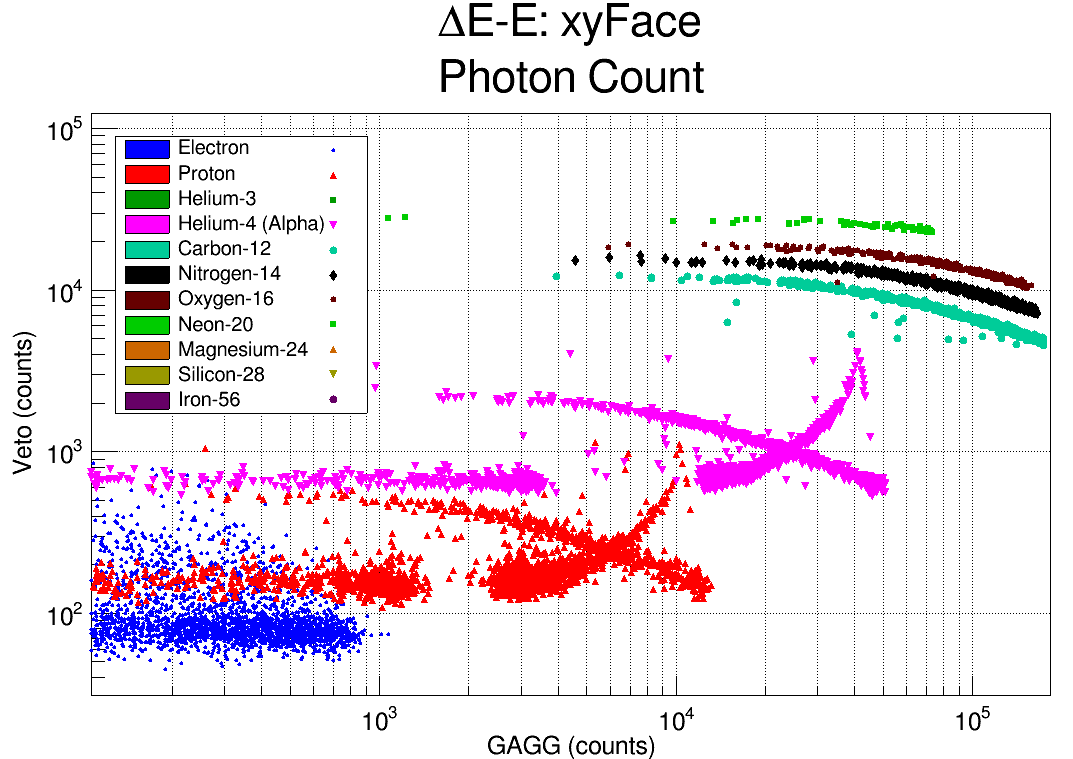}
    \caption{Veto vs GAGG photon count particle plot for the xyFace scenario.}
    \label{xyFacePhotonP}
\end{figure}

\begin{figure}[h!]
    \centering
    \includegraphics[width=0.88\textwidth]{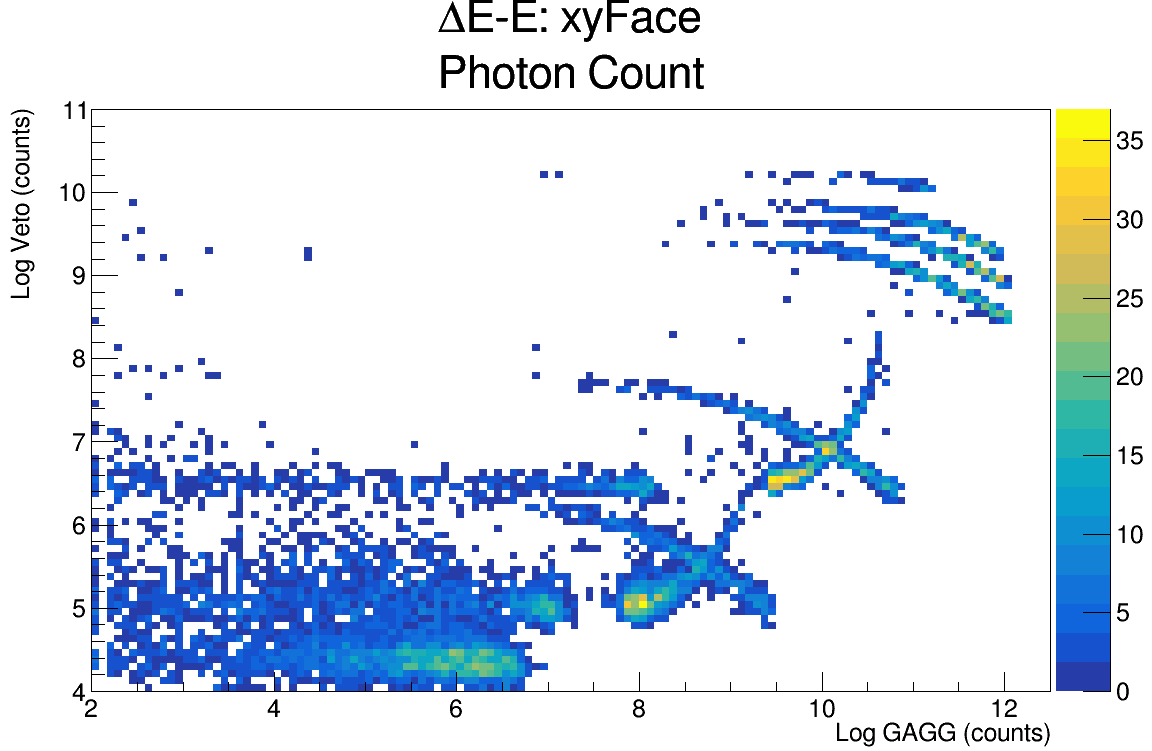}
    \caption{Veto vs GAGG photon count histogram for the xyFace scenario.}
    \label{xyFacePhotonH}
\end{figure}

Moving on to the xyFace photon count study shown in Figure~\ref{xyFacePhotonP} and Figure~\ref{xyFacePhotonH}, we see our first major deviation from the pencil beam scenario. There is a discontinuity in the proton and alpha spectra that is not seen in the pencil beam or xyFace energy deposition plots. However, because this does not affect the different particle regions, $\Delta$E-E particle discrimination still properly separates the particles in this scenario.

\subsubsection{Impingement of the z+Face}
\begin{figure}[h!]
    \centering
    \includegraphics[width=\textwidth]{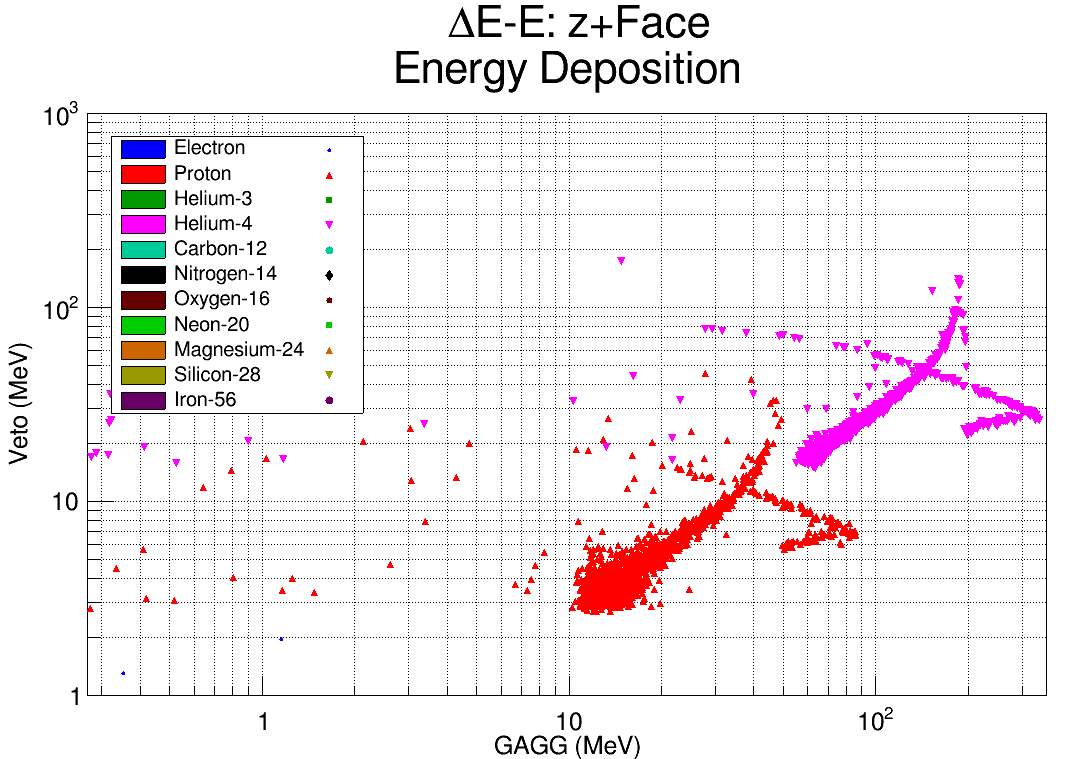}
    \caption{Veto vs GAGG energy deposition particle plot for the z+Face scenario.}
    \label{z+FaceEdepP}
\end{figure}

\begin{figure}[h!]
    \centering
    \includegraphics[height=10cm]{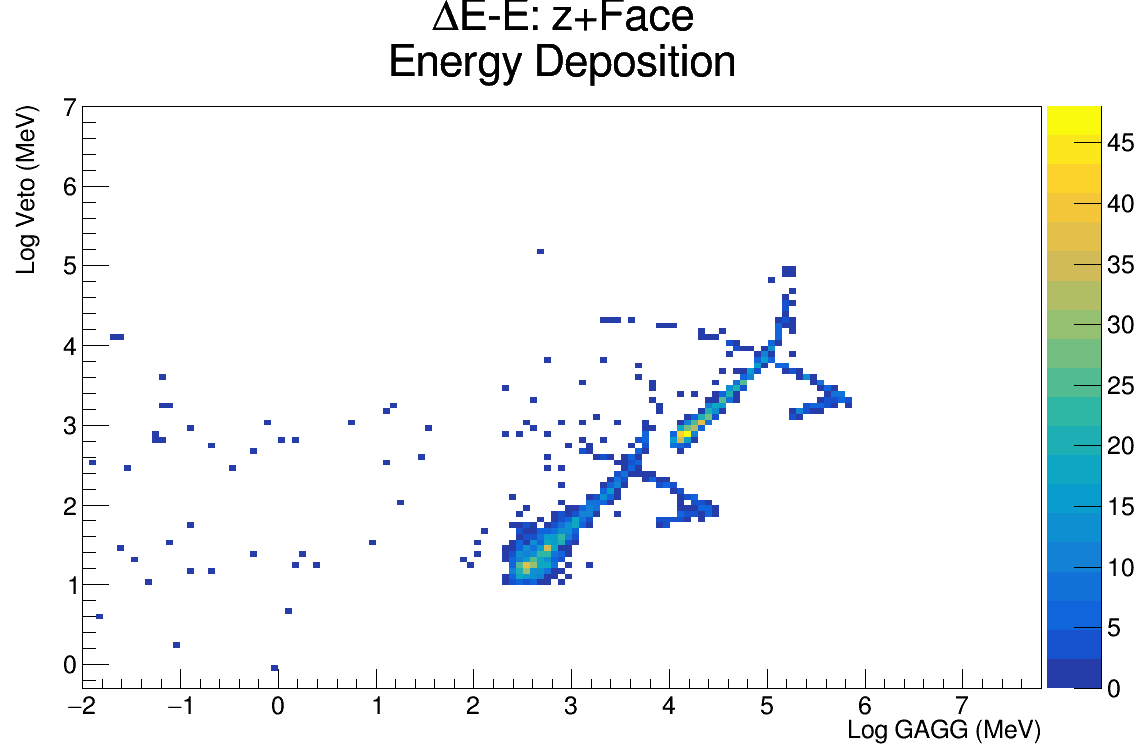}
    \caption{Veto vs GAGG energy deposition histogram for the z+Face scenario.}
    \label{z+FaceEdepH}
\end{figure}

$\tab$The z+Face differs significantly from the previous two simulation scenarios, as shown in Figure~\ref{z+FaceEdepP} and Figure~\ref{z+FaceEdepH}. This is due to the difference in shielding that is provided by the detector in this scenario. Each particle must get through the entire Veto PMT, before it even has a chance of depositing energy into the Veto or GAGG. If particles are coming from the positive Z face, it can be inferred that we should only expect protons and alpha particles to deposit energy in the Veto and GAGG. This is shown in Figure~\ref{z+FaceEdepP} by the absence of blue, dark green, teal, black, brown, light green, orange, gold, or purple dots on the plot.

It is also seen that the general shape of the $\Delta$E-E curve varies from the pencil beam and xyFace scenario. 

\begin{figure}[h!]
    \centering
    \includegraphics[height=10cm]{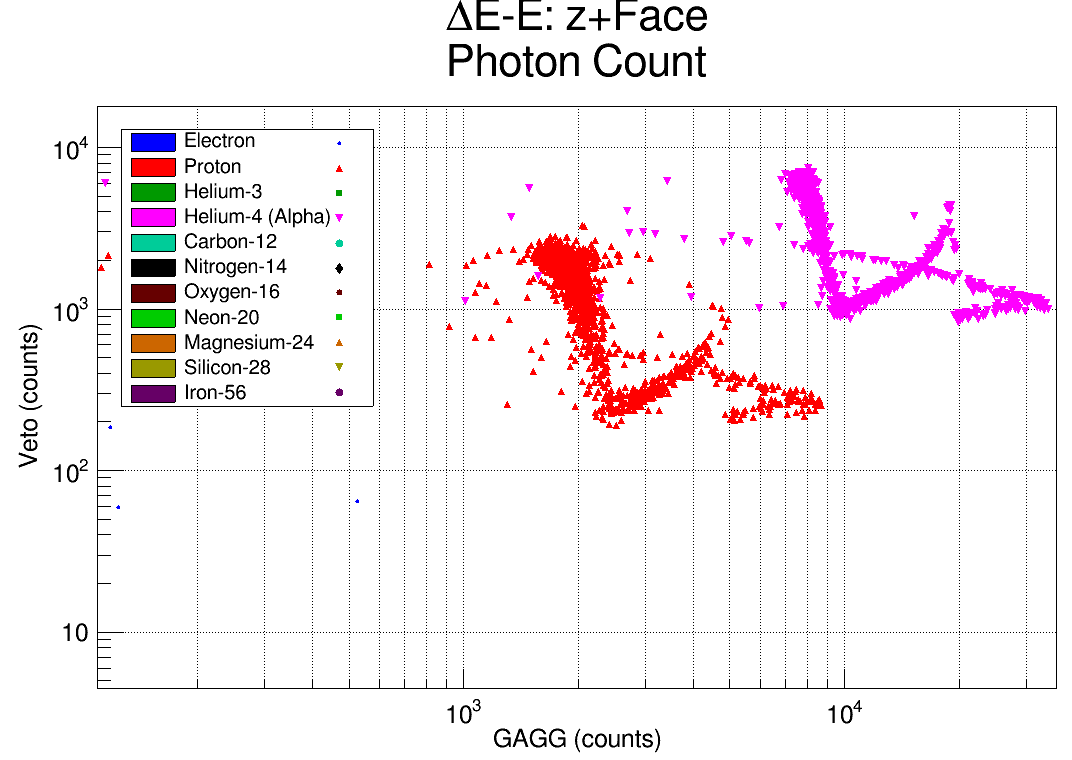}
    \caption{Veto vs GAGG photon count particle plot for the z+Face scenario.}
    \label{z+FacePhotonP}
\end{figure}

\begin{figure}[h!]
    \centering
    \includegraphics[height=10cm]{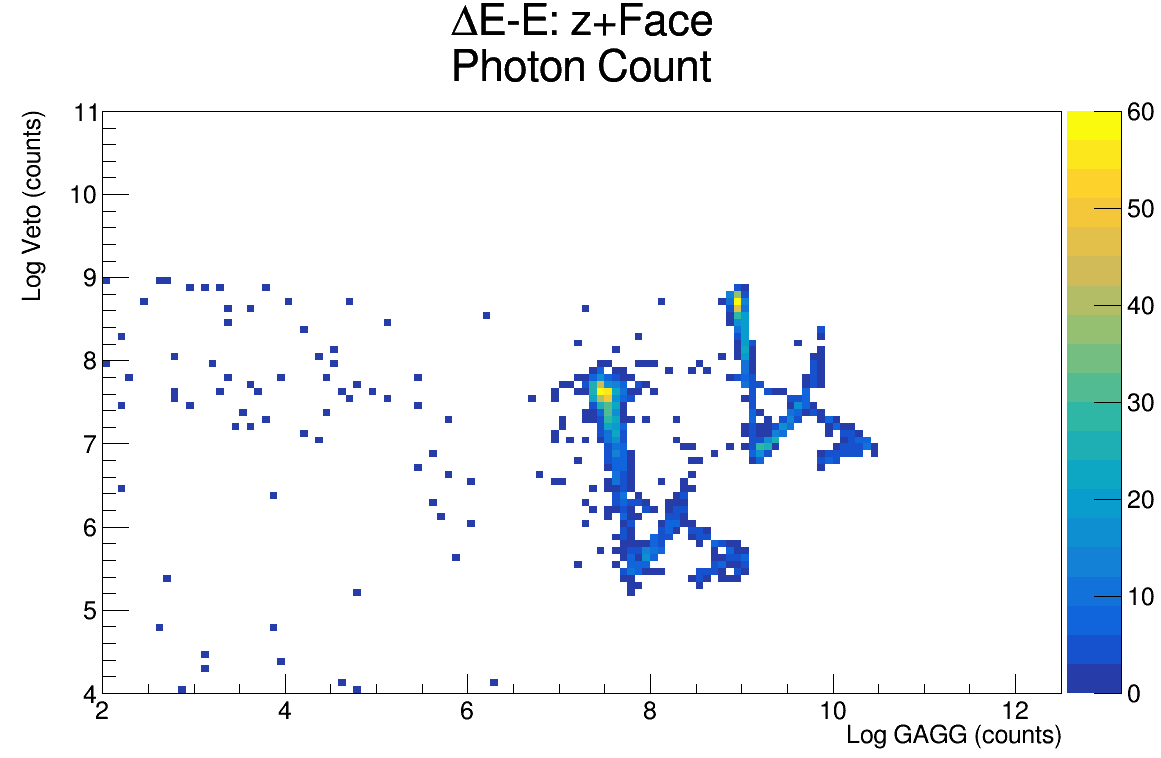}
    \caption{Veto vs GAGG photon count histogram for the z+Face scenario.}
    \label{z+FacePhotonH}
\end{figure}

The photon count plots in Figure~\ref{z+FacePhotonP} and Figure~\ref{z+FacePhotonH} again have a different shape than what was found in the previous scenarios. The different shape follows the same trend that the energy deposition plots show, with further differences. The photon count plots have a V shape that seems to pop out. 

The difference in shape between the energy deposit and photon count plots again do not take away from the particle separation. $\Delta$E-E is still a valid particle discrimination method for this scenario.

\subsubsection{Impingement of the z-Face}
$\tab$The z-Face again differs significantly from the pencil beam and xyFace scenarios, but has notable similarities to the results of the z+Face scenario. The z-Face scenario energy deposition plots have very similar shape, with the same particles reaching the Veto and GAGG as the z+Face scenario. Likewise, the photon count plots of the z-Face have the same V that shows up in the z+Face scenario. However, these photon count plots vary from the z+Face to the lower right of the V shape.

This scenario provides a different angle on $\Delta$E-E. This scenario shows what plots will look like when the particle must deposit energy in volumes other than just the Veto and GAGG. When the particle gets through the shielding, it deposits energy first into the Veto. Instead of going straight to the GAGG, the particle must deposit energy in the electronics, SiPMs, and the quartz crystal window, before finally depositing energy into the GAGG. 

\begin{figure}[h!]
    \centering
    \includegraphics[height=10cm]{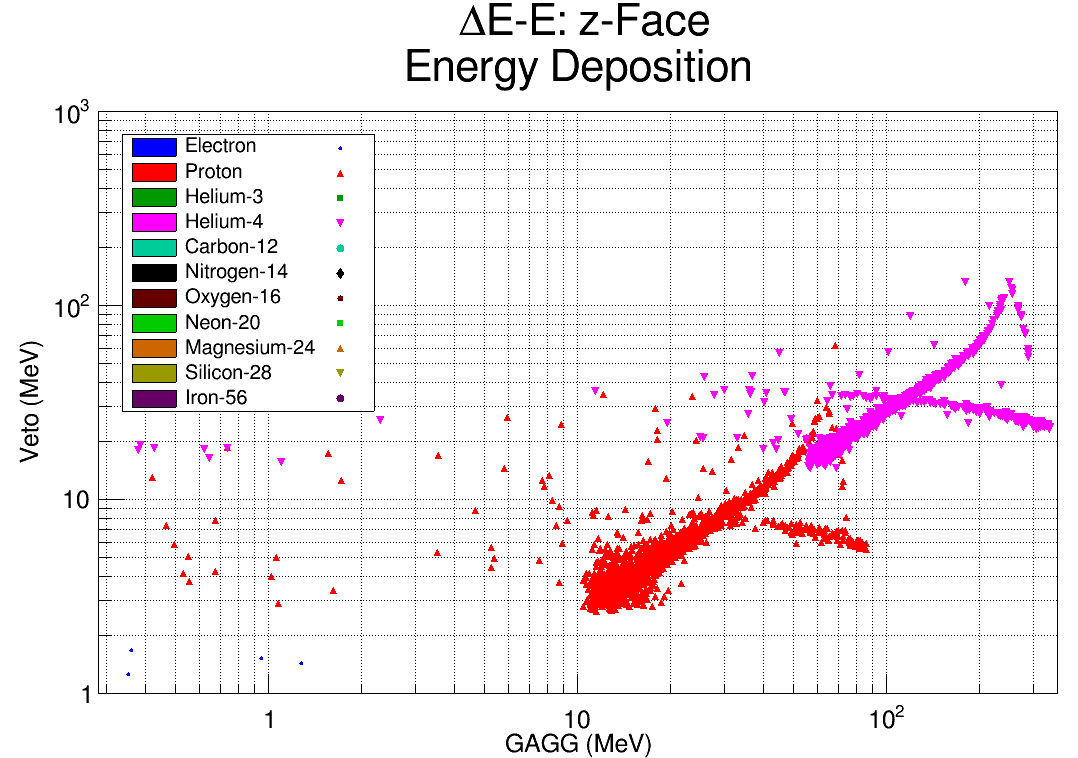}
    \caption{Veto vs GAGG energy deposition particle plot for the z-Face scenario.}
    \label{z-FaceEdepP}
\end{figure}

\begin{figure}[h!]
    \centering
    \includegraphics[height=10cm]{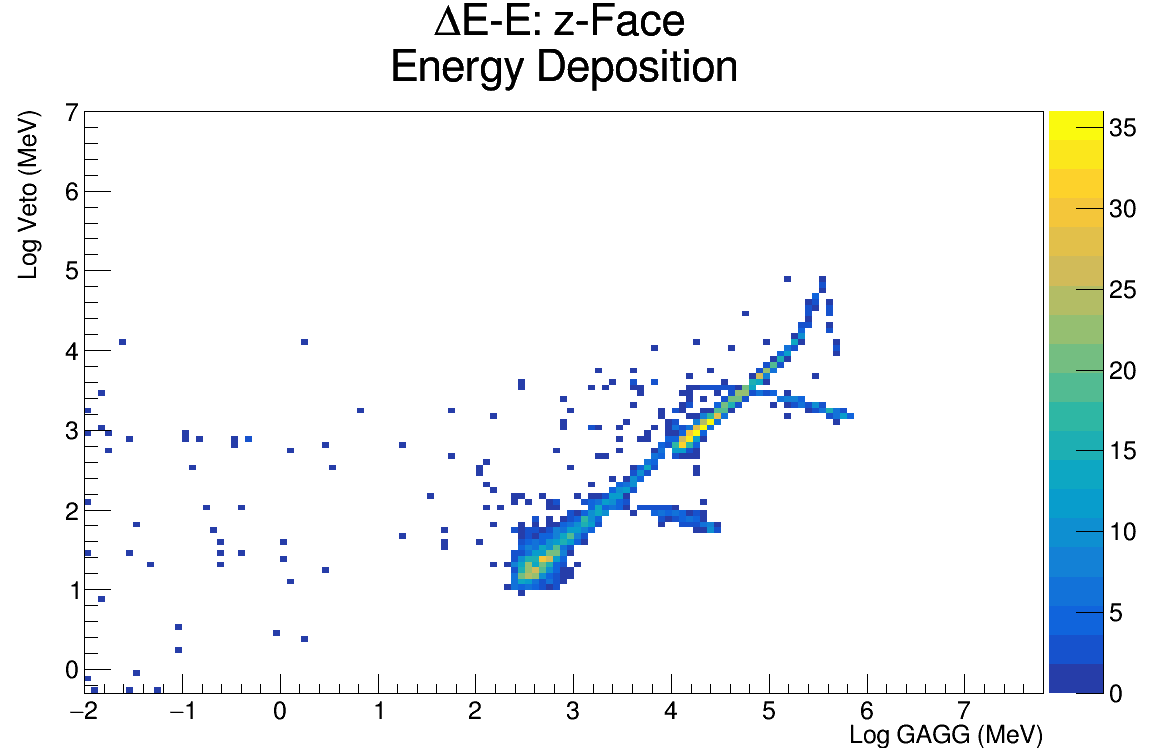}
    \caption{Veto vs GAGG energy deposition histogram for the z-Face scenario.}
    \label{z-FaceEdepH}
\end{figure}

\begin{figure}[h!]
    \centering
    \includegraphics[height=10cm]{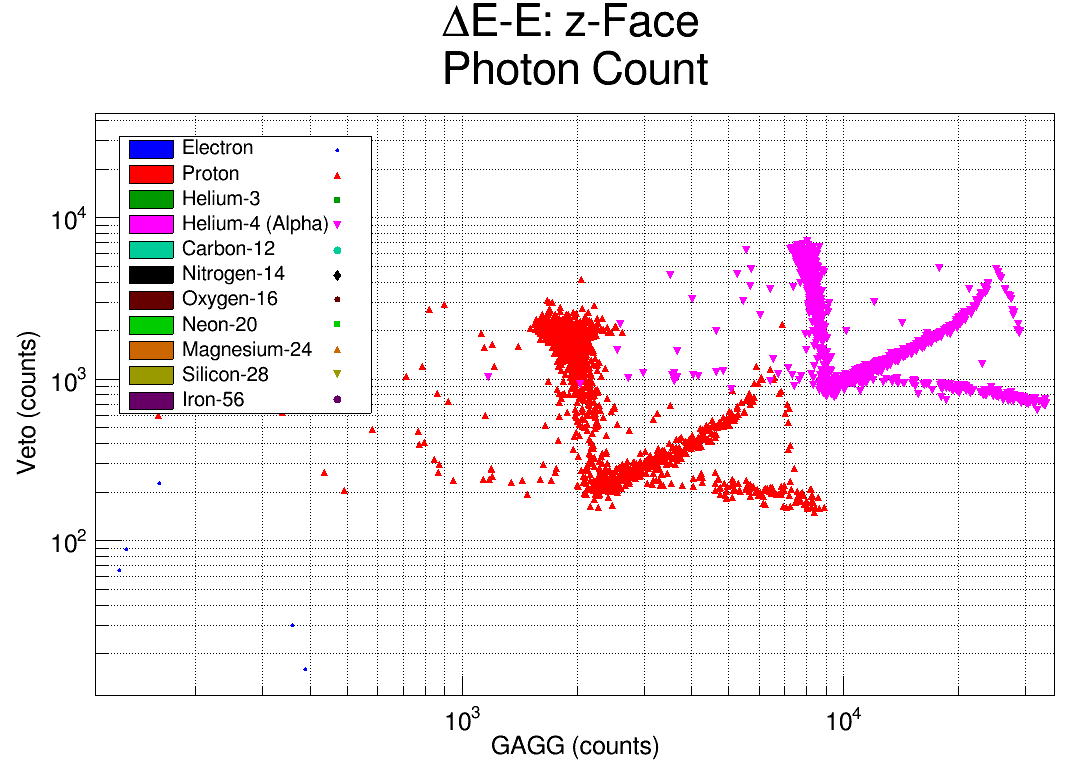}
    \caption{Veto vs GAGG photon count particle plot for the z-Face scenario.}
    \label{z-FacePhotonP}
\end{figure}

\begin{figure}[h!]
    \centering
    \includegraphics[height=10cm]{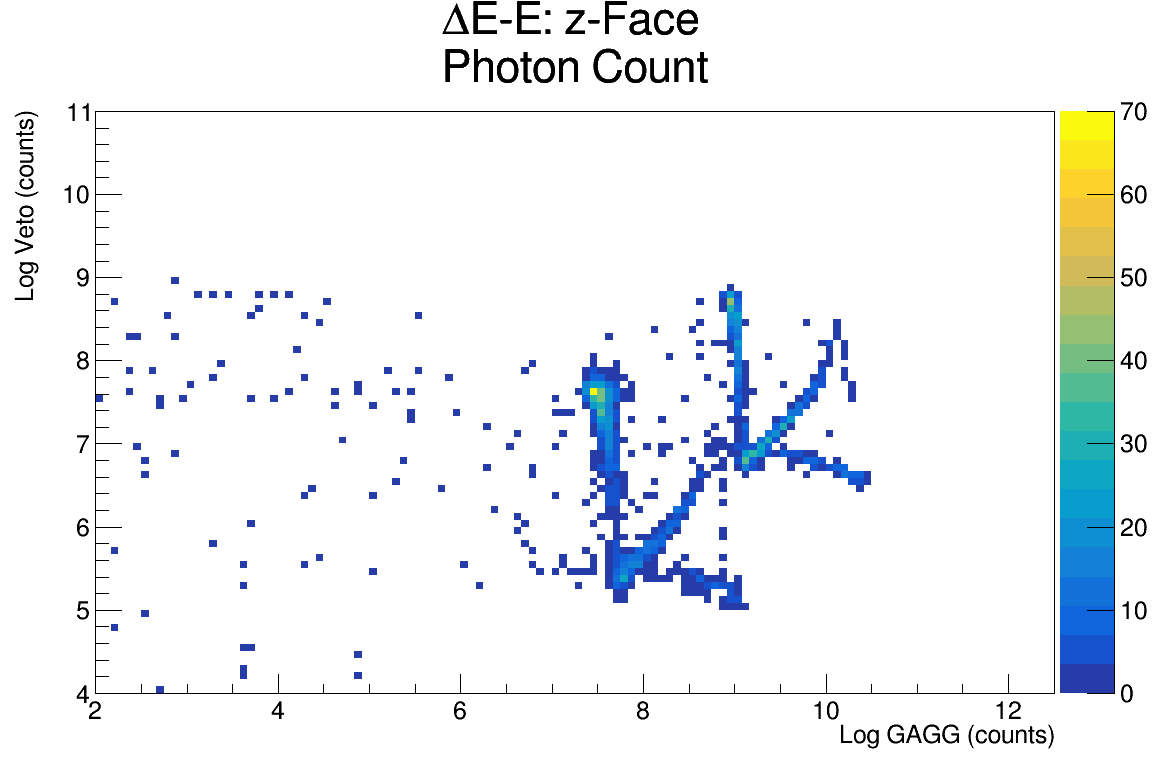}
    \caption{Veto vs GAGG photon count histogram for the z-Face scenario.}
    \label{z-FacePhotonH}
\end{figure}

Due to this different geometry, this scenario does not at all follow the methods of the general $\Delta$E-E telescope. However, as can be seen from Figures ~\ref{z-FaceEdepP} -~\ref{z-FacePhotonH}, the different particle regions continue to separate out very nicely. Therefore, for the z-Face scenario, particle discrimination via $\Delta$E-E will continue to be a viable method.

Note the overlap of the top right of the proton region and bottom left of the alpha region. While this overlap does exist, the histogram in Figure~\ref{z-FaceEdepH} shows that this is insignificant to the differentiation of particles.

\subsection{Isotropic Simulations}
$\tab$The isotropic simulations correspond to the simulation scenarios in which all the particles that are incident on the detector are isotropic in their flight path. As stated before, these scenarios are called iso, hemisphereIsoZ+Face, and hemisphereIsoZ-Face. This will best describe solar wind measurements while in orbit.

\subsubsection{Spherical Isotropic Impingement}
$\tab$The results from the isotropic scenarios will be a far cry from the results found in the pencil beam and Face scenarios. This fact will hold true for this and both hemisphere isotropic scenarios. The purpose of looking at the iso scenario first is to show what can be expected from combining both the +Z and -Z hemispheres. After which, we will look at each hemisphere separately.

\begin{figure}[h!]
    \centering
    \includegraphics[height=9.6cm]{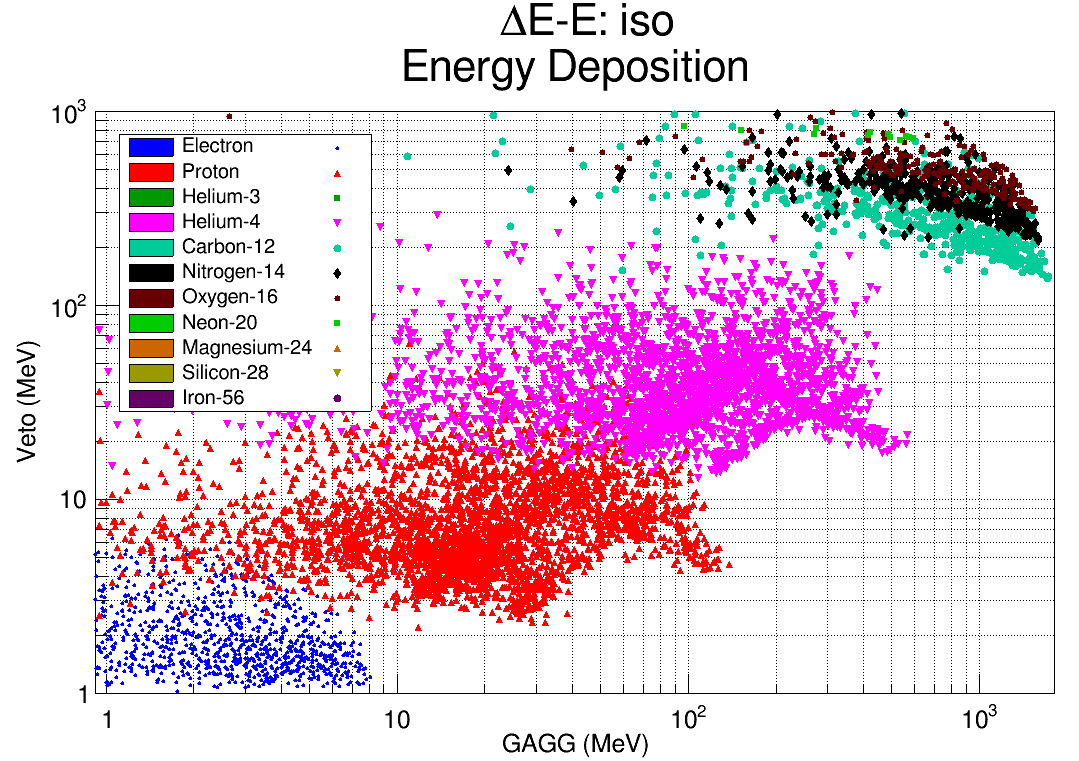}
    \caption{Veto vs GAGG energy deposition particle plot for the iso scenario.}
    \label{isoEdepP}
\end{figure}

\begin{figure}[h!]
    \centering
    \includegraphics[height=9.6cm]{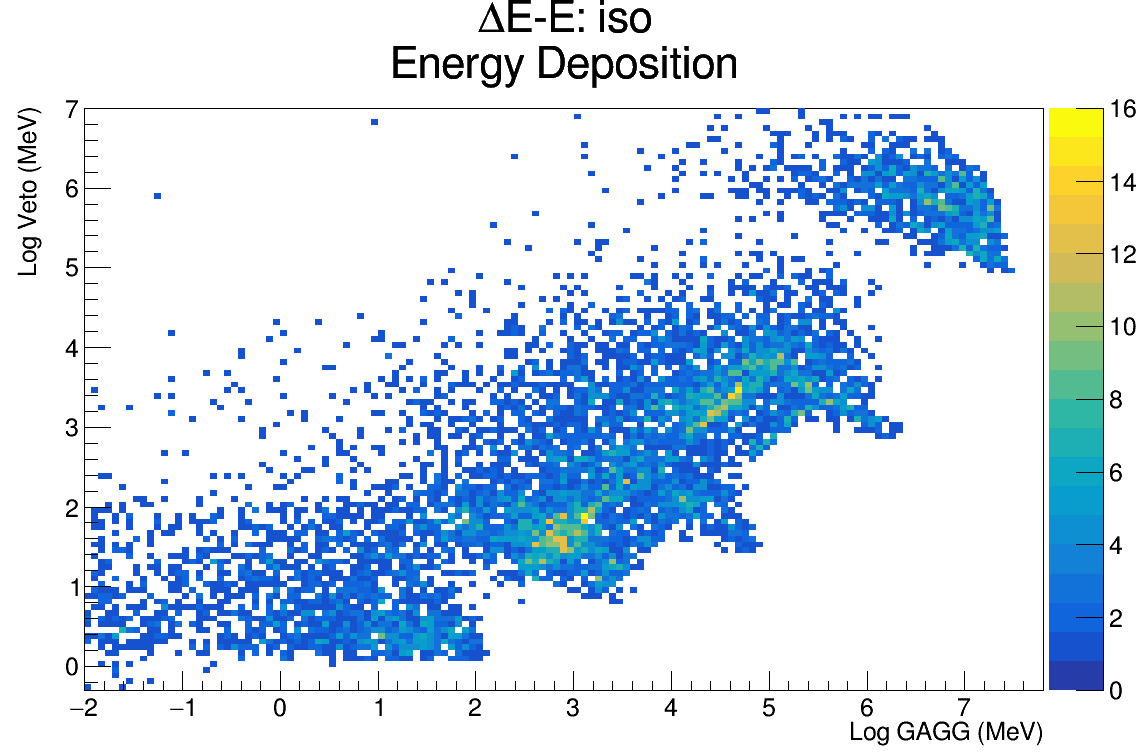}
    \caption{Veto vs GAGG energy deposition histogram for the iso scenario.}
    \label{isoEdepH}
\end{figure}

\begin{figure}[h!]
    \centering
    \includegraphics[height=10cm]{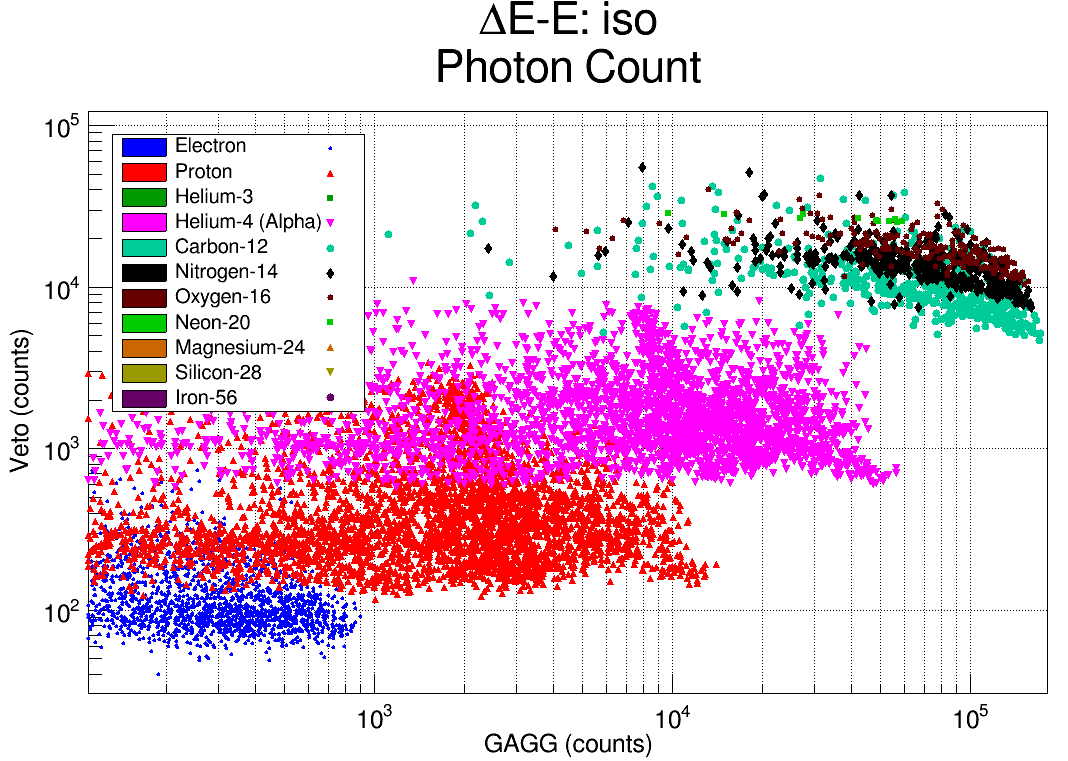}
    \caption{Veto vs GAGG photon count particle plot for the iso scenario.}
    \label{isoPhotonP}
\end{figure}

\begin{figure}[h!]
    \centering
    \includegraphics[height=10cm]{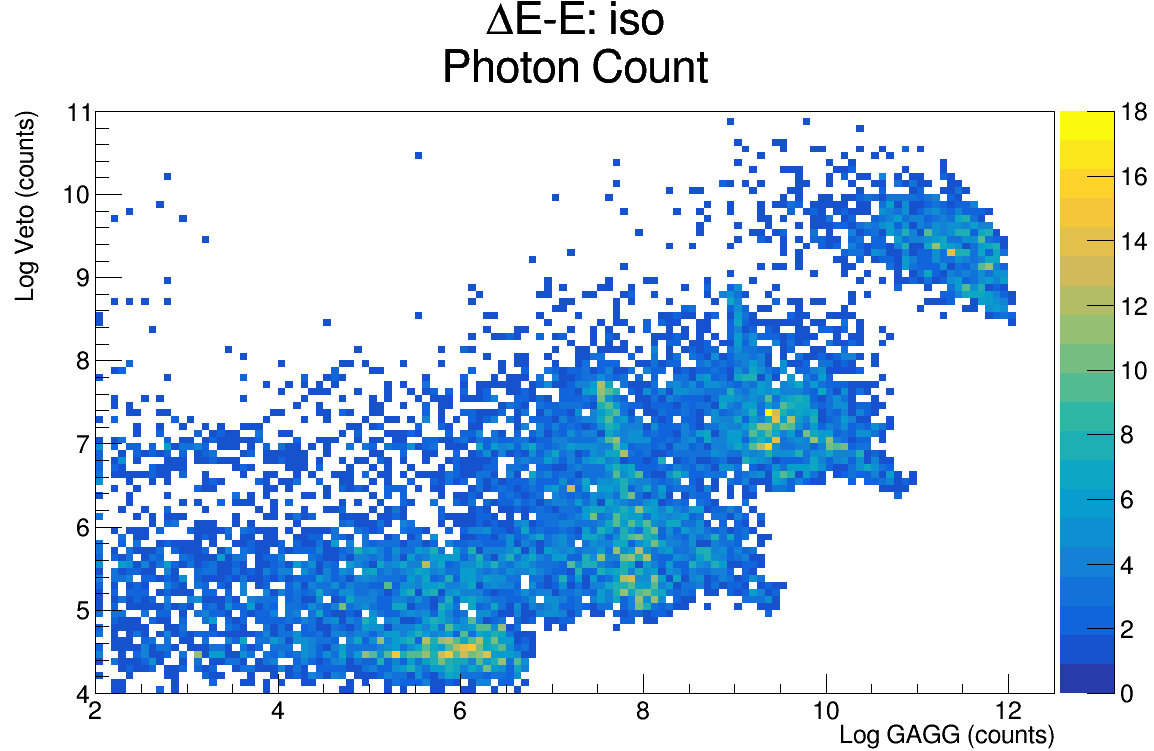}
    \caption{Veto vs GAGG photon count histogram for the iso scenario.}
    \label{isoPhotonH}
\end{figure}

As can be seen in Figure~\ref{isoEdepP} and Figure~\ref{isoPhotonP}, $^3$He, $^{24}$Mg, $^{28}$Si, and $^{56}$Fe are not expected to get through the shielding (similar to the pencil beam and xyFace scenarios). Because the particle flight path is isotropic, none of the particles will likely be coming from the exact same direction. This is reflected in each of Figures ~\ref{isoEdepP} -~\ref{isoPhotonH}, with no smooth curves to be found. This being the case, electrons, protons, and alpha particles seem to have relatively distinct regions within the plot. This would indicate that for lighter particles, particle discrimination via $\Delta$E-E is still a valid method. Recall that electrons, protons, and alpha particles are by far the most abundant particles in solar energetic particle events.

However, $^{12}$C, $^{14}$N, $^{16}$O, and $^{20}$Ne seem to get lumped together in the same region. This would indicate that for heavier ions, particle discrimination via $\Delta$E-E is not a valid method.

The photon count study once again shows slight differences to the energy deposition study. In the case of the iso scenario, the proton and alpha particle regions take on what I would call a dolphin shape, as can be seen in Figure~\ref{isoPhotonP}. This will explain the dense vertical section in the proton region that is found in Figure~\ref{isoPhotonH}. This region has some overlap with the alpha particle region, which is why it is so dense on the histogram. This overlap can be seen in Figure~\ref{isoPhotonP}.

\subsubsection{Hemispherical Isotropic Impingement of the z+Face}
$\tab$Not all that much can be said for the hemisphereIsoZ+Face scenario, as the results are very similar to those found in the iso scenario. The electron, proton, and alpha particle regions are separated, similar to how they were in the iso scenario, and the $^{12}$C, $^{14}$N, $^{16}$O, and $^{20}$Ne regions seem to be combined, also just as they were in the iso scenario. This result is to be expected given the isotropic nature of the particle flight path. 

The results of the photon count study are again very similar to the iso scenario. Once again the electron, proton, and alpha particle regions separate nicely, while the $^{12}$C, $^{14}$N, $^{16}$O, and $^{20}$Ne regions again combine into a single region.

\begin{figure}[h!]
    \centering
    \includegraphics[width=0.89\textwidth]{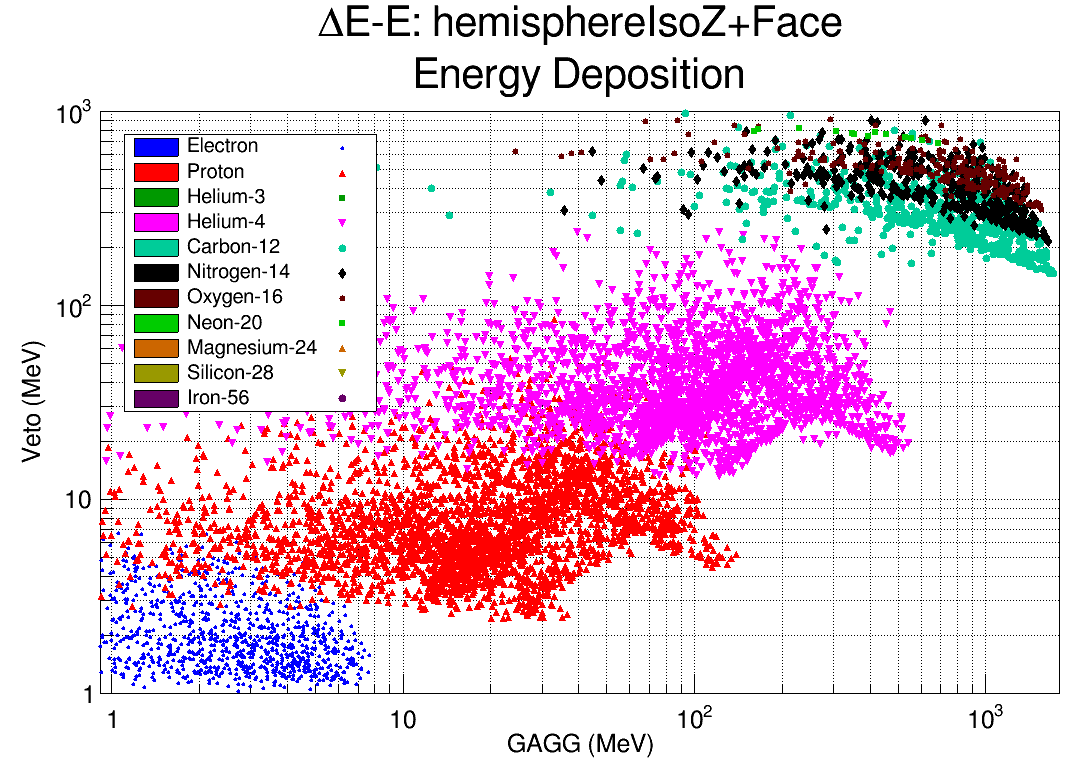}
    \caption{Veto vs GAGG energy deposition particle plot for hemisphereIsoZ+Face.}
    \label{hemisphereIsoZ+FaceEdepP}
\end{figure}

Note the similarities in the shape of the different regions for both energy deposition and photon count between the iso scenario and hemisphereIsoZ+Face scenario.

\begin{figure}[h!]
    \centering
    \includegraphics[width=0.89\textwidth]{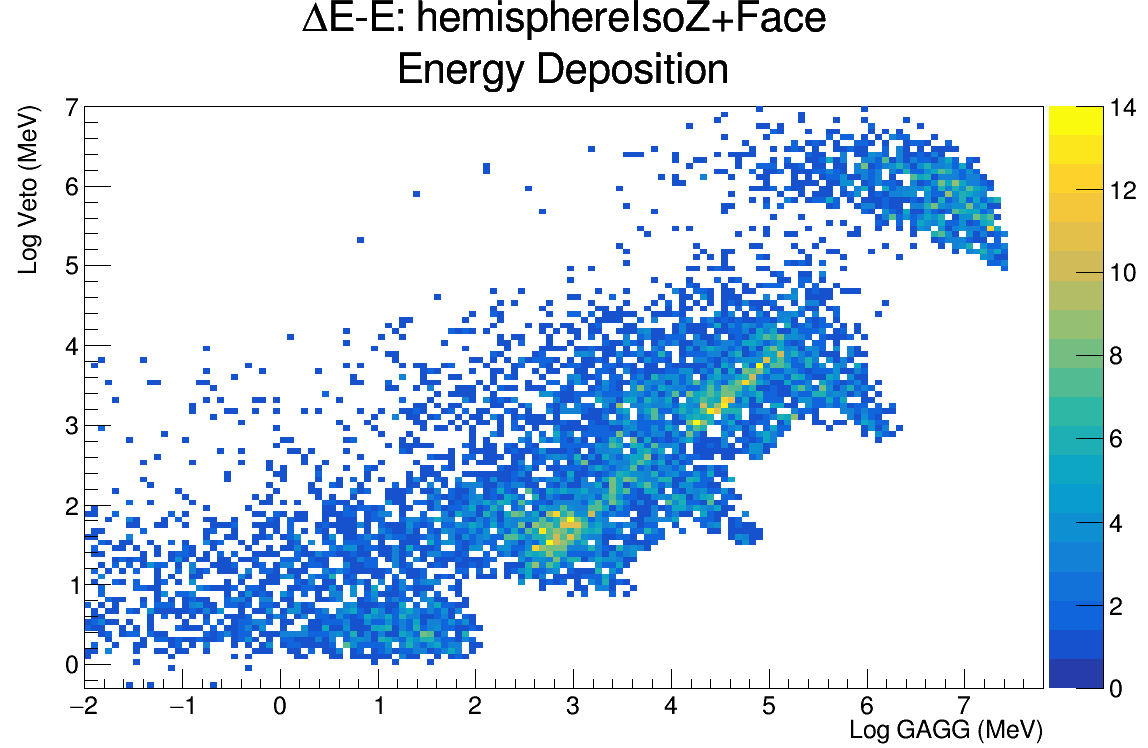}
    \caption{Veto vs GAGG energy deposition histogram for hemisphereIsoZ+Face.}
    \label{hemisphereIsoZ+FaceEdepH}
\end{figure}

\begin{figure}[h!]
    \centering
    \includegraphics[width=0.89\textwidth]{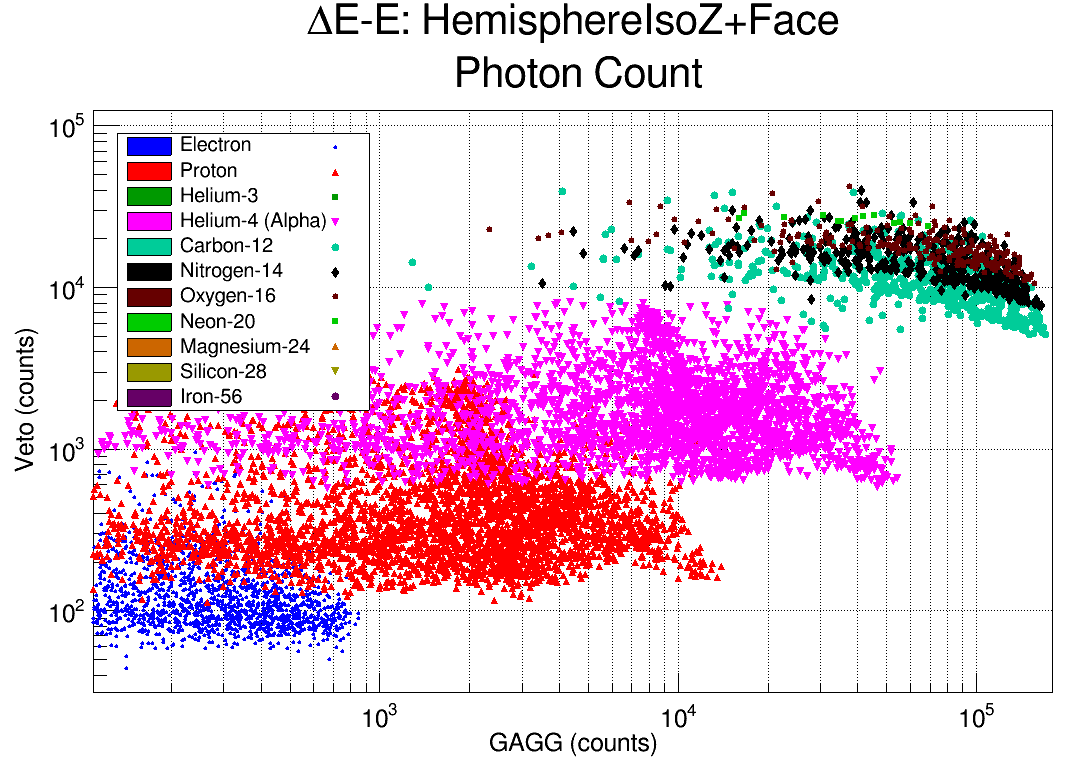}
    \caption{Veto vs GAGG photon count particle plot for hemisphereIsoZ+Face.}
    \label{hemisphereIsoZ+FacePhotonP}
\end{figure}

\begin{figure}[h!]
    \centering
    \includegraphics[width=0.89\textwidth]{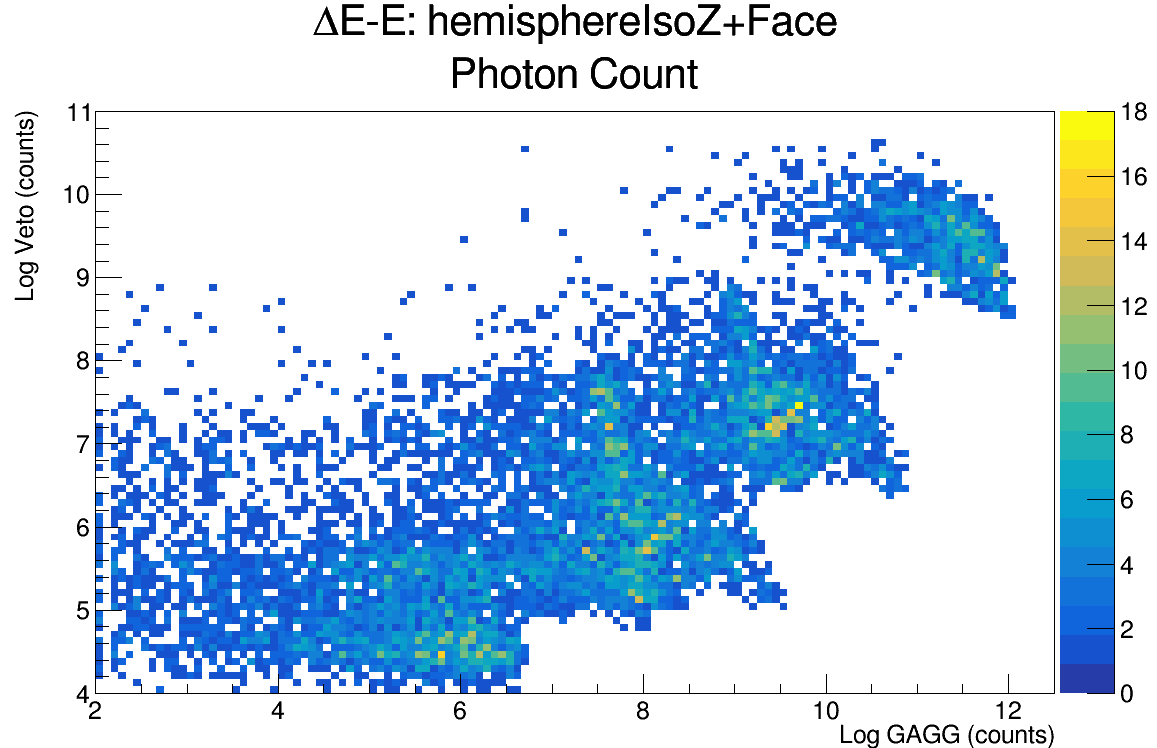}
    \caption{Veto vs GAGG photon count histogram for hemisphereIsoZ+Face.}
    \label{hemisphereIsoZ+FacePhotonH}
\end{figure}

\subsubsection{Hemispherical Isotropic Impingement of the z-Face}
$\tab$The results of the hemisphereIsoZ-Face simulation are similar enough to the iso and hemisphereIsoZ+Face scenarios that comments will not be necessary.

\begin{figure}[h!]
    \centering
    \includegraphics[width=0.79\textwidth]{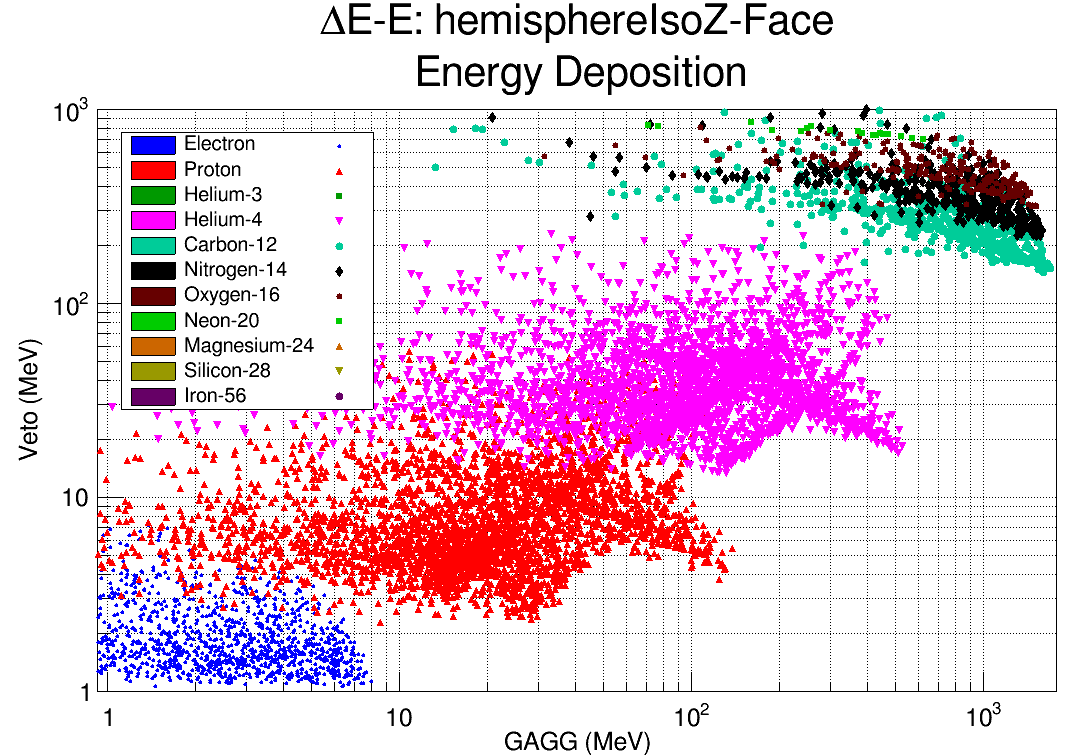}
    \caption{Veto vs GAGG energy deposition particle plot for hemisphereIsoZ-Face.}
    \label{hemisphereIsoZ-FaceEdepP}
\end{figure}

\begin{figure}[h!]
    \centering
    \includegraphics[width=0.79\textwidth]{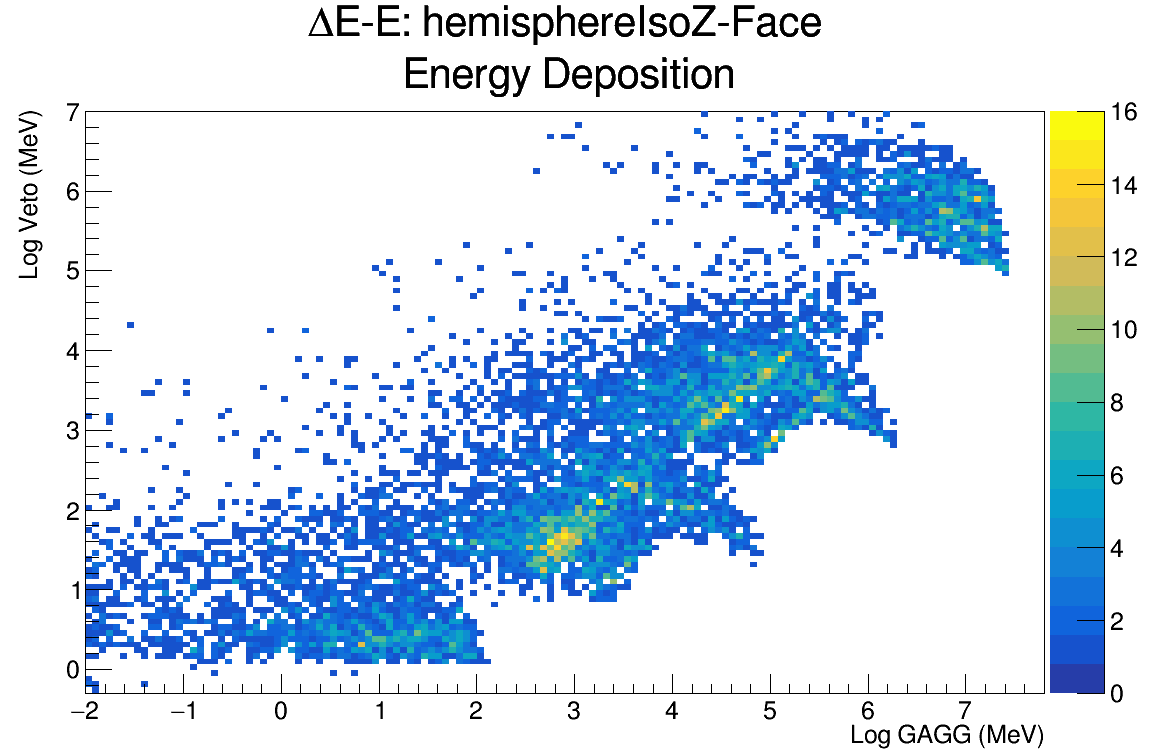}
    \caption{Veto vs GAGG energy deposition histogram for hemisphereIsoZ-Face.}
    \label{hemisphereIsoZ-FaceEdepH}
\end{figure}

\begin{figure}[h!]
    \centering
    \includegraphics[width=0.89\textwidth]{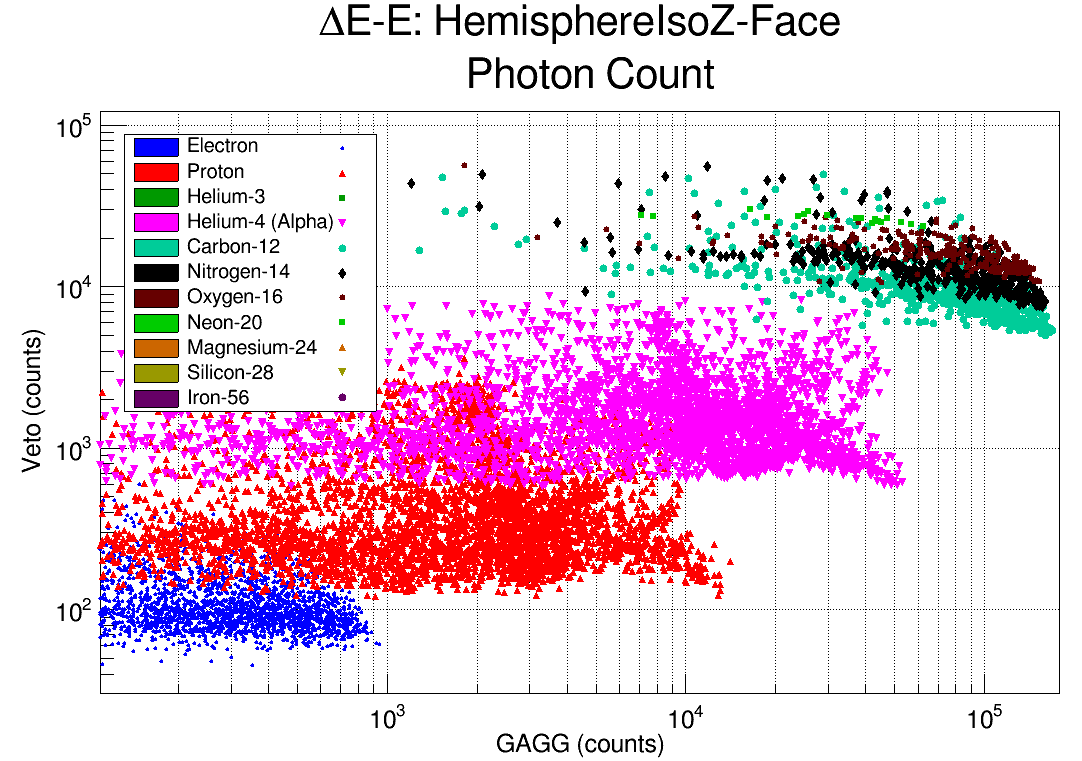}
    \caption{Veto vs GAGG photon count particle plot for hemisphereIsoZ-Face.}
    \label{hemisphereIsoZ-FacePhotonP}
\end{figure}

\begin{figure}[h!]
    \centering
    \includegraphics[width=0.89\textwidth]{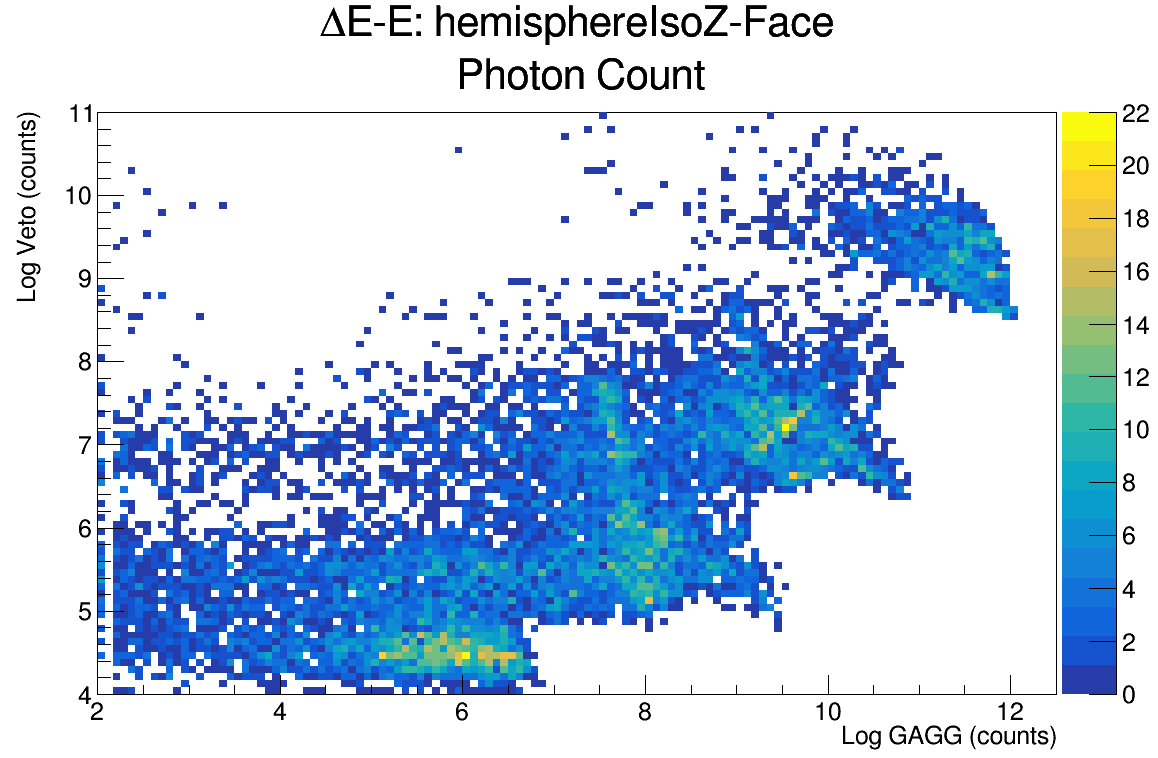}
    \caption{Veto vs GAGG photon count histogram for hemisphereIsoZ-Face.}
    \label{hemisphereIsoZ-FacePhotonH}
\end{figure}

\section{Conclusion}
$\tab$The presented simulation was constructed in GEANT4 using a model of the CubeSat detector. The simulation data came from energy deposition measurements and the photon count process within GEANT4. The simulated particles were determined to be the most abundant particles in solar energetic particle events, with their energies corresponding to the most probable range. 

The presented results demonstrate the differences between the CubeSat detector and the more general $\Delta$E-E telescope. For the pencil beam and Face simulation scenarios, the CubeSat detector performs on par with $\Delta$E-E telescopes, with differences occurring due to the geometry of the CubeSat detector. In the isotropic scenarios, the heavy ions get bunched together and are not fully distinguishable. However, a majority of simulation scenarios show that electrons, protons, and alpha particles separate into distinct regions of the plot, which indicates particle discrimination via $\Delta$E-E is a valid method for electrons, protons, and alpha particles.

Future work will use the region of good selection of electrons, protons and alpha particles to then determine the correction factors due to acceptance and shielding distortions such that the measured particle energy spectrum can be turned into the actual particle energy spectrum for solar wind. These acceptance corrections will be deduced mostly from Monte-Carlo simulations but some individual points can be confirmed with comparison to particle accelerator test beam data of the measured detectors and veto arrays response signals.

\section*{Acknowledgments}
$\tab$I would like to thank Dr. Nick Solomey for giving me the opportunity to work on the \textnu SOL team, and for his direction on where to focus my efforts. I would also like to thank Jonathan Folkerts for introducing me to the C++ language and GEANT4, without which I would not have been able to complete this research. I am grateful to Brian Doty for allowing me to use the model of the CubeSat detector that he created. I am also thankful to Brookes Hartsock, Tyler Nolan, and Dr. Holger Meyer for all of the help they have extended to me.

Special thanks to Dr. Dan Smith, a 2024 Watkins Summer Research Fellow, whose work with the \textnu SOL project validated the results of this simulation.

The NASA Jump Start program in Kansas is funded by NASA grant 80NSSC20M0109.

\end{document}